\documentclass[aps,pra,twocolumn,groupedaddress,showpacs,floatfix,superscriptaddress,longbibliography]{revtex4-2}
\usepackage{epsfig}
\usepackage{amsmath,amssymb}
\usepackage{graphicx}
\usepackage[dvipsnames,usenames]{color}
\usepackage[normalem]{ulem}
\usepackage{soul} 
\usepackage[plainpages=false,pdfpagelabels,colorlinks=true,linkcolor=red,urlcolor=blue,citecolor=blue,pdftitle={Title},pdfauthor={},pdfdisplaydoctitle=true,pdfduplex=DuplexFlipLongEdge]{hyperref}

\graphicspath{{./plots/}}
\begin{document}

\title{Photoinduced enhancement of superconductivity in the plaquette Hubbard model}
\author{Yuxi Zhang}
\affiliation{Department of Physics, University of California, Davis, CA 95616,USA}
\author{Rubem Mondaini} 
\affiliation{Beijing Computational Science Research Center, Beijing 100084, China}
\author{Richard T. Scalettar}
\affiliation{Department of Physics, University of California, Davis, CA 95616,USA}
\date{\today}

\begin{abstract}
Real-time dynamics techniques have proven increasingly useful in understanding strongly correlated systems both theoretically and experimentally. By employing unbiased time-resolved exact diagonalization, we study pump dynamics in the two-dimensional plaquette Hubbard model, where distinct hopping integrals $t_h$ and $t_h^\prime$ are present within and between plaquettes. In the intermediate coupling regime, a significant enhancement of $d$-wave superconductivity is observed and compared with that obtained by simple examination of expectation values with the eigenstates of the Hamiltonian. Our work provides further understanding of superconductivity in the Hubbard model, extends the description of the pairing amplitude to the frequency-anisotropy plane, and offers a promising approach for experimentally engineering emergent out-of-equilibrium states.
\end{abstract}

\maketitle

\section{Introduction}

Out-of-equilibrium dynamics and photoinduced properties have been attracting intense attention recently. In a single-particle setting, periodic drives provide an additional ``Floquet dimension", and can result in phenomena such as engineered Chern insulating behavior and  Wannier-Stark localization~\cite{oka19}. These techniques have also allowed considerable new insight into strongly correlated systems, and many-body phases induced by either electron-electron or electron-phonon interactions~\cite{wang2017,wang2018t}, including studies of photoinduced $\eta$-pairing~\cite{kaneko2019,kaneko2020}, bond-order wave (BOW) physics~\cite{shao2019}, dynamical topological engineering~\cite{mciver20, shao21}, and especially superconductivity~\cite{cavalleri2018,peronaci2015,wang2016,sentef2017,wang2018,wang2021,tang2021}. A focus has been on gaining new insight into cuprate systems, and the origin of pairing in high-$T_c$ materials~\cite{fausti2011,kaiser2014,nicoletti2014,cremin2019}.
 
One mechanism for the enhancement of superconductivity is the use of photoexcitation to transfer charge between orbitals, thereby shifting the doping away from 1/8 filling where the ``stripe anomaly" leads to a suppressed $T_c$. Here exact diagonalization studies have clarified whether the emergence of superconductivity via photoexcitation seen experimentally results from phonon or electronic mechanisms~\cite{wang2018}. Indeed,  superconductivity can also be developed out of the 1/8 filling phase  through photoexcitation of the Cu-O stretch mode in the plane of the  stripes~\cite{cavalleri2018}. This coupling to vibrational modes offers advantages in that the lower energy photons have less likelihood of creating high temperature electronic distributions and heating the lattice.
 
In calculations similar to the study to be described here, a theoretical understanding of the enhancement of pairing is provided through the lifting of the degeneracy between superconducting and charge density wave (CDW) ground states in the attractive Hubbard Hamiltonian~\cite{sentef2017}. In the Hubbard-Holstein model, inducing $d$-wave pairing has been shown to be especially effective at the quantum boundary between the CDW and spin density wave (SDW) ground state phases~\cite{wang2018}. Theoretical work within a BCS model has also explored the role of the symmetry of the pair wave function. The existence of nodes in a $d$-wave phase allows for a more rapid decay, making the dynamics substantially faster than the $s$-wave channel~\cite{peronaci2015}.

In parallel with these developments, theoretical studies have shown that inhomogeneities such as stripe-like or plaquette decorations of two-dimensional square lattices could enhance $d$-wave pairing in equilibrium~\cite{tsai2006,yao2007,tsai2008,doluweera2008,yao2010,smith13,ying2014}. This is an especially rich area of inquiry given the complex   entwining of spin and charge textures with pairing~\cite{kivelson03,berg09,vojta09,huang18,agterberg20,tranquada20}. Specifically, under some circumstances inhomogeneities clearly interfere with superconductivity, while in others they appear to play a role in its formation and stability.
  
In this paper we {\it combine} these two themes of out-of-equilibrium dynamics and spatial inhomogeneity. In particular, we examine the effect of pumping on a geometry built up from attractive $d$-wave binding centers formed by $2\times 2$ plaquettes. Our goal is to understand the interplay of these two effects which are already known {\it individually} to result in enhanced superconductivity. Our work will therefore examine the possibility that a greater signal of pairing might be thereby engineered, 
a prospect made particularly intriguing by the recent experimental emulation of fermionic models in  $2\times 2$ plaquettes in platforms of quantum dots~\cite{Dehollain2020}.
  
In the next section we introduce the plaquette Hubbard model and our computational methodology. We then examine the degree of $d$-wave pairing in the different low energy eigenstates. A central result here is a `phase diagram' which shows the degree of enhanced superconductivity in the plane of hopping anisotropy $t_h^\prime/t_h$ and excitation energy $E_{\alpha}-E_0$. We next examine time evolution, with the objective of employing a quasi-periodic drive of an appropriate frequency $\omega_0$ to target those optimal $E_{\alpha}-E_0$. The end result is a second `phase diagram' which indicates the achievable enhancement in the plane of $\omega_0$ and drive amplitude $A_0$. Here we focus on the $t_h^\prime/t_h$ which gives optimal pairing in the analysis of the eigenstates. We conclude the paper with a summary of results and a discussion of the general significance of our findings.
  
\section{Model and Methodology}

\begin{figure}[t!]
\includegraphics[width=.9\columnwidth]{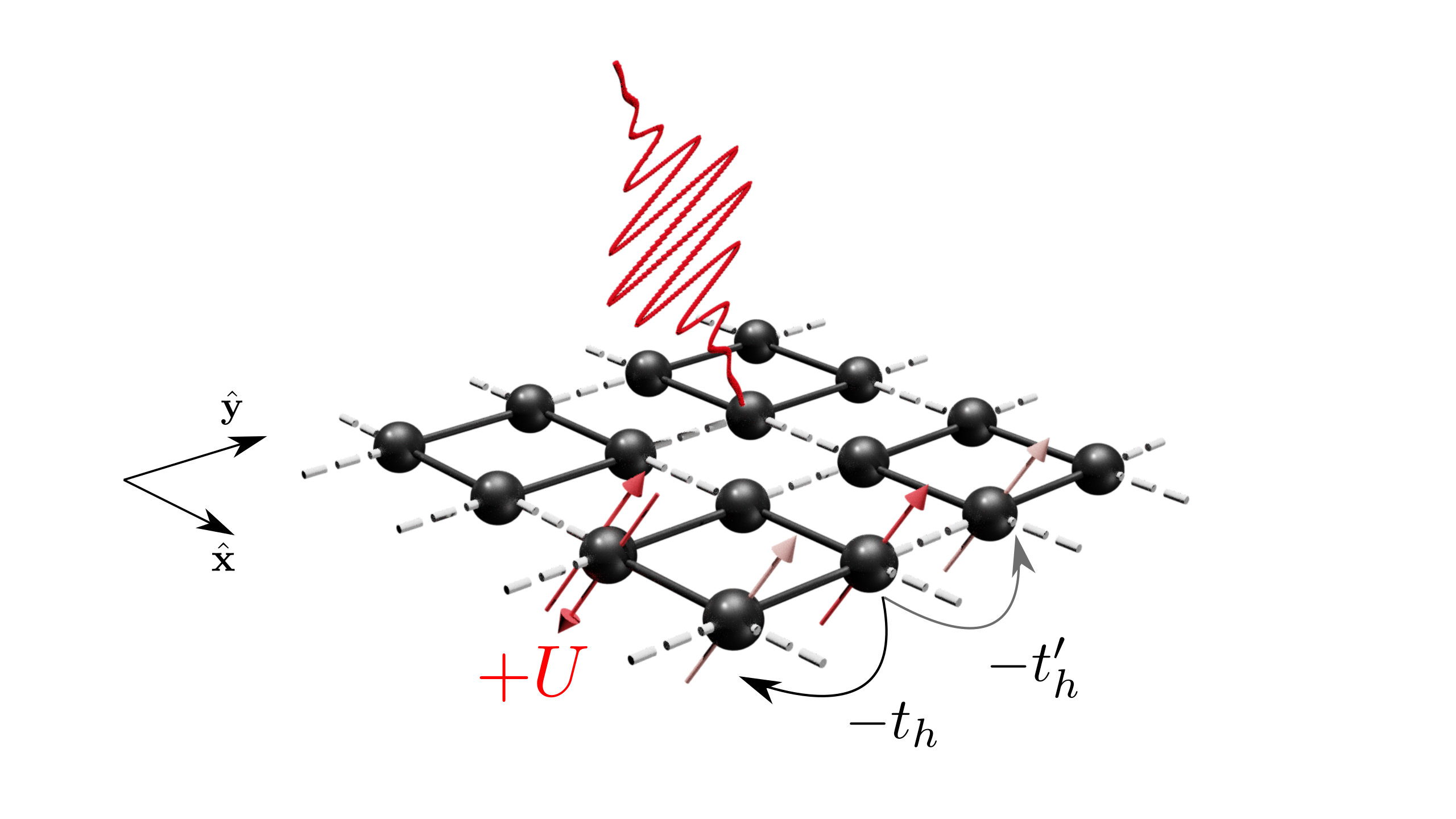} %
\caption{Schematic illustration of the plaquette Hubbard model on a $4\times4$ lattice with PBCs, with relevant terms in the Hamiltonian annotated. Dark (light) colored bonds represent the intra-(inter-) plaquette hopping $t_h(t_h')$.
}
\label{fig:lattice} 
\end{figure}

The two dimensional Hubbard model is described by 
\begin{align}
\mathcal{\hat H} = & - \sum_{\langle \mathbf{i}, \mathbf{j}
  \rangle, \sigma} \big(t_{\mathbf{i},\mathbf{j}}^{\phantom{\dagger}}
\, \hat c^{\dagger}_{\mathbf{i} \sigma} \hat
c^{\phantom{\dagger}}_{\mathbf{j} \sigma} + {\rm H.c.} \big) 
+ U \sum_{\mathbf{i}} \hat
n_{\mathbf{i}, \uparrow} \hat n_{\mathbf{i}, \downarrow} \,\, ,
\label{eq:Hamiltonian}
\end{align}
where $\hat c^{\phantom{\dagger}}_{\mathbf{i} \sigma}
  (\hat c^{\dagger}_{\mathbf{i} \sigma})$ is the annihilation (creation) operator for an electron on site $\mathbf{i}$ with spin  $\sigma=\uparrow, \downarrow$ and $\hat n_{\mathbf{i},\sigma}^{\phantom{\dagger}}=
 \hat c^{\dagger}_{\mathbf{i},\sigma} \hat
 c^{\phantom{\dagger}}_{\mathbf{i},\sigma}$ 
 is the number operator. $t_{\mathbf{i},\mathbf{j}}=t_h^{\phantom{'}}(t_h')$ gives the hopping integral within (between) plaquettes, and on-site electron-electron interaction is tuned via $U$. We set $t_h=1$ as the energy scale, and the time $t$ is measured in terms of $t_h^{-1}$ with $\hbar=1$. Throughout this work, we use an $N =4\times4$ lattice with periodic boundary conditions (PBC), as illustrated in Fig.~\ref{fig:lattice}.

The system is driven by a time-dependent external field, which is introduced in the Hamiltonian via the Peierls’ substitution 
\begin{align}
t_{\mathbf{i},\mathbf{j}}^{\phantom{\dagger}}
\, \hat c^{\dagger}_{\mathbf{i} \sigma} \hat
c^{\phantom{\dagger}}_{\mathbf{j} \sigma} + {\rm H.c.} \rightarrow t_{\mathbf{i},\mathbf{j}}^{\phantom{\dagger}} \,e^{i\mathbf{A}(t) \cdot (\mathbf{r}_\mathbf{i}-\mathbf{r}_\mathbf{j})}
\, \hat c^{\dagger}_{\mathbf{i} \sigma} \hat
c^{\phantom{\dagger}}_{\mathbf{j} \sigma} + {\rm H.c.} \,\,,
\end{align}
where $\mathbf{A}(t)$ is the spatially uniform vector potential. We choose $\mathbf{A}(t)$, mimicking an ultrafast photoirradiation, of the form
\begin{align}
\mathbf{A}(t)=A_0 e^{-(t-t_0)^2/2 t_d^2}\, \cos[\omega_0(t - t_0)+\varphi]\,\mathbf{e}_{\rm pol}\,\, ,
\label{eq:vec_pot}
\end{align}
i.e., an oscillatory Gaussian centered at $t_0$ and width $t_d$, with amplitude $A_0$, frequency $\omega_0$ and polarization direction $\mathbf{e}_{pol}=(\hat {\mathbf x} + \hat {\mathbf y})/\sqrt{2}$. Finally, $\varphi$ is a time-phase of the pump pulse, allowing the possibility of phase-averaging that eliminates particular fast coherent oscillations tied to interaction magnitudes~\cite{wang2017}. When applicable, results of the dynamics will employ an average over 10 equidistant values of $\varphi$ in the range $[0,2\pi)$. In what follows, we set $t_0=100\,t_h^{-1}$ and $t_d=25\,t_h^{-1}$, with maximum time chosen as $t_{\rm max}=200\,t_h^{-1}$, which is well after the pump still exhibits a relevant amplitude.

Simulations are carried out with the time-dependent Lanczos method~\cite{Manmana2005,Prelovsek}, where the time-evolution is obtained via $|\Psi(t+dt)\rangle=e^{-i\mathcal{ \hat H}(t)dt}|\Psi(t)\rangle$, with $dt$ a sufficiently small time step. The initial state, $|\Psi(t\to -\infty)\rangle$, is the ground state of \eqref{eq:Hamiltonian}, with $|\Psi(t)\rangle$ and associated measurements being evaluated at each time step.

\section{Equilibrium Results} \label{sec:equil}
To map out where an out-of-equilibrium protocol would induce enhanced superconductivity, we start by examining the equilibrium properties of the plaquette Hubbard model across a range of $(t_h'/t_h^{\phantom{'}}, U, n_e)$ parameters. Some of these have already been explored at finite temperatures by means of quantum Monte Carlo simulations~\cite{ying2014} or at the ground-state~\cite{tsai2008} in lattice sizes as in Fig.~\ref{fig:lattice}. Here, we go beyond that and investigate different equal-time correlators computed over the low-lying eigenspectrum of \eqref{eq:Hamiltonian} $[\hat {\cal H} |\alpha\rangle = E_\alpha |\alpha\rangle]$. Among them, we calculate the pairing structure factor at different $\gamma$-channels 
\begin{align}
P_{\gamma} = \frac{1}{N} \sum_{{\bf i},{\bf dr}}  \langle \hat \Delta_{\bf i}^{(\gamma)\phantom{\dagger}} \hat \Delta_{\bf i + dr}^{(\gamma)\dagger}\rangle \,\, ,
\end{align}
where $\hat \Delta_{\bf i}^{(\gamma)\phantom{\dagger}} = (\hat c_{{\bf i + \hat x},\sigma} + f(\gamma) \hat c_{{\bf i + \hat y},\sigma})\hat c_{{\bf i},\overline \sigma}$, with $f(\gamma) = -1(+1)$ for $d$-wave(extended $s$-wave, $s^*$) pairing. In a similar fashion, the \textit{staggered} spin and charge correlations are probed as
\begin{align}
S_{x} = \frac{1}{N} \sum_{\bf i,dr} \langle(-1)^{dx+dy} \hat O^{\dagger}_{\bf i} \hat O^{\phantom{\dagger}}_{\bf i+dr}\rangle \,\, ,
\end{align}
where $\hat O_{\bf i}=\hat n_{\bf i\uparrow} \pm \hat n_{\bf i\uparrow}$, and the $+$ sign defines charge ($x=$ CDW) and the $-$ sign spin ($x=$ SDW) density wave order. 

Over the large set of parameters available, we focus the investigation on both the doped ($\rho \equiv n_e/N = 0.875$) and undoped cases ($\rho =1$); the latter described in the Appendix \ref{app:half_fill}. Due to geometric constraints in the lattice size we tackle, a period-8 magnetic stripe formation is not supported~\cite{Zheng2017,Huang2017,huang18,Ponsioen19}. The ground state properties of $\hat {\cal H}$, originally reported in Ref.~\onlinecite{tsai2008}, indicated that a maximum binding energy occurs at intermediate plaquette hopping energies ($t^\prime_h/t_h^{\phantom{'}} \simeq 0.5$), and large repulsive interactions $U\simeq8\,t_h$. This serves as a paradigm to our investigation, in which we are interested in the set of parameters that maximize the expectation values of pairing correlators over states beyond that of the ground state $|0\rangle$. 

\begin{figure}[t!]
\includegraphics[width=1\columnwidth]{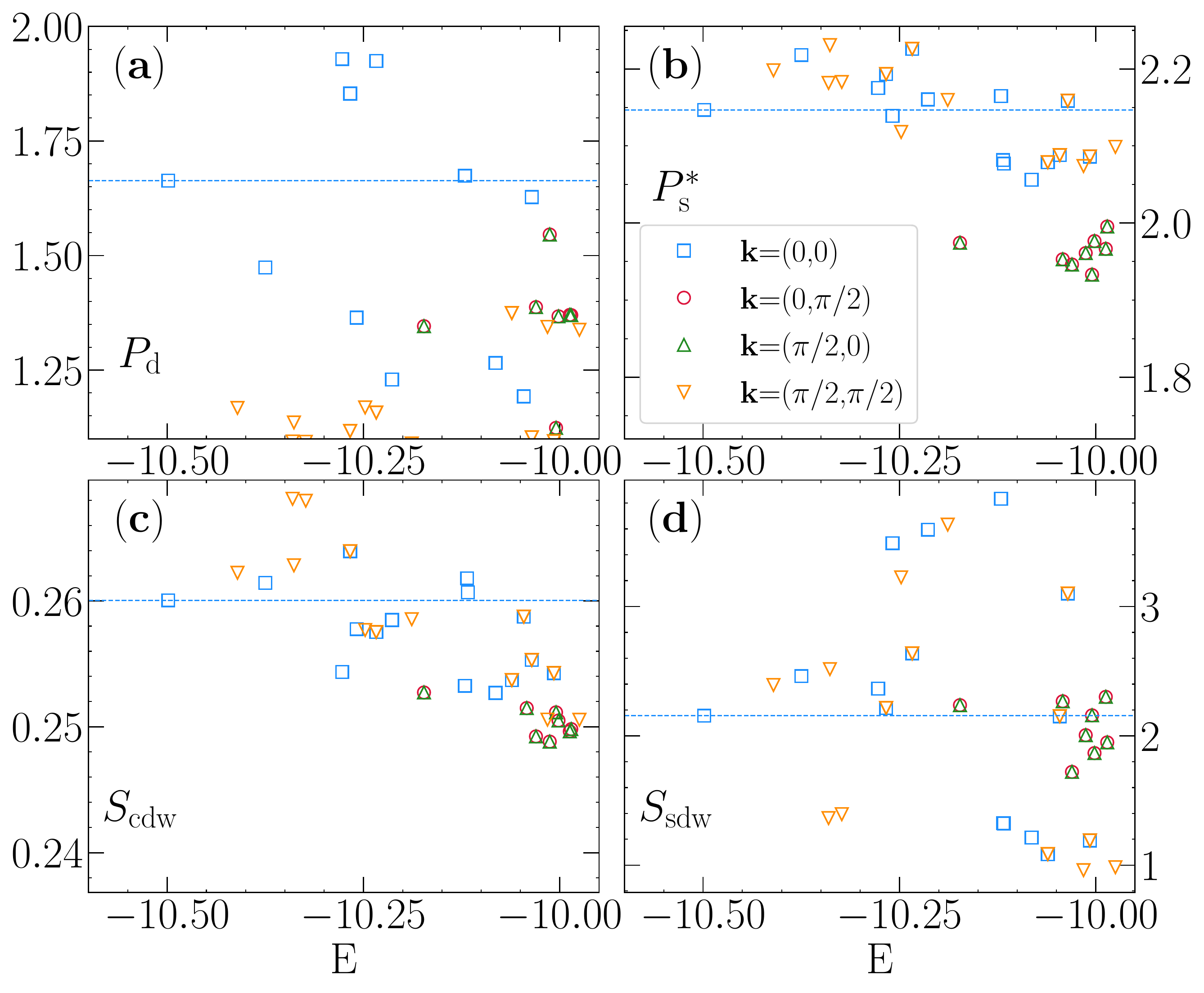} %
\caption{(a) $d$-wave, (b) extended $s$-wave, (c) CDW and (d) SDW structure factor, in the ground state (dashed lines) and the first few excited states. Different colors and symbols represent different quasi-momentum sectors. Parameters used are $t_h'/t_h^{\phantom{'}}=0.8$, $U=8\,t_h$ and $n_e=14$. We show the different momentum sectors ${\bf k}$ for completeness, since for this filling the ground state always resides at ${\bf k} = (0,0)$. Each sector displays various eigenstates $|\alpha\rangle$, some of which may display degeneracies owing to other point group symmetries not being resolved, 
e.g.~mirror symmetries.
}
\label{fig:static} 
\end{figure}

\begin{figure}[t!]
\includegraphics[width=0.84\columnwidth]{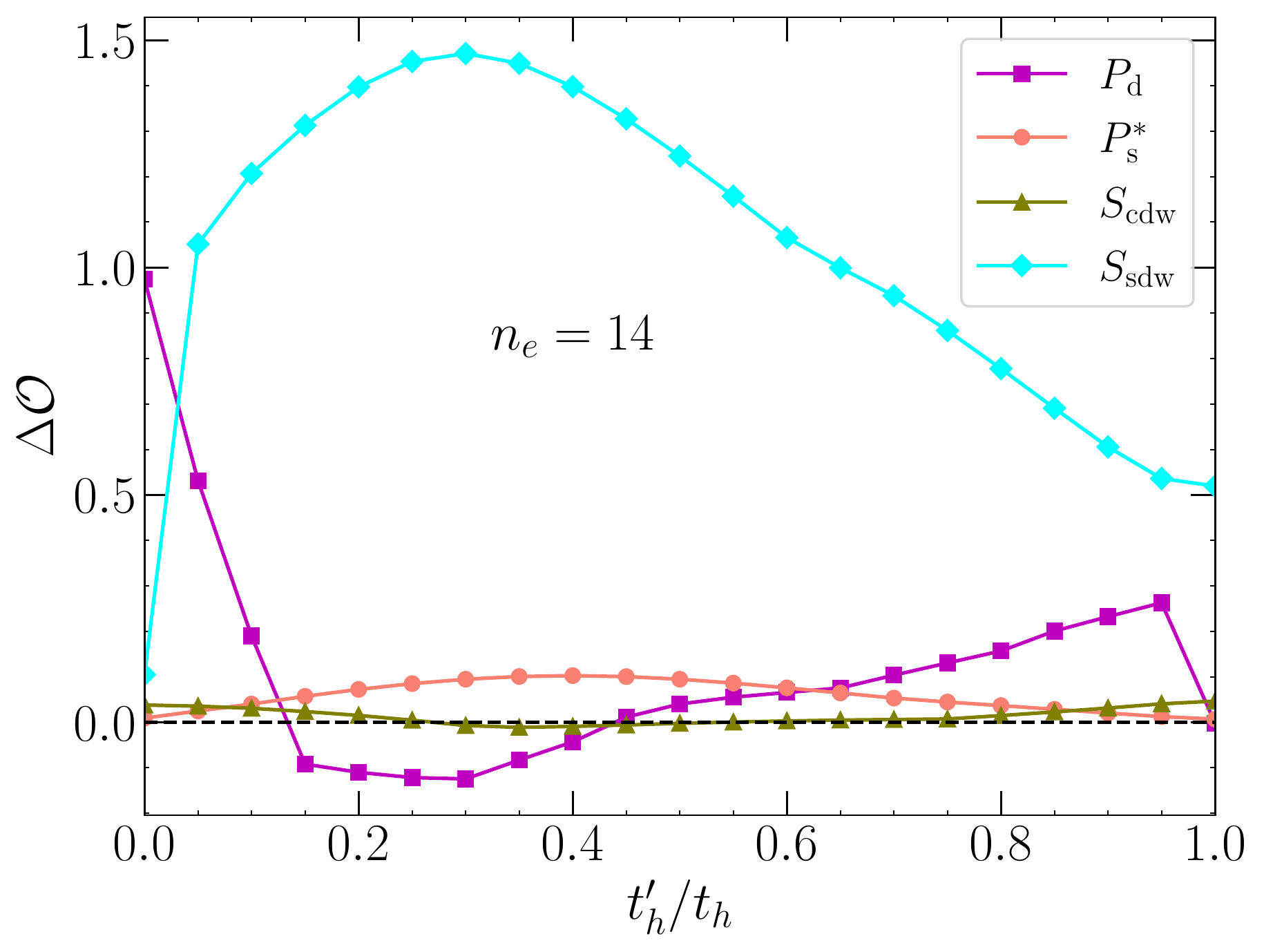} %
\caption{Relative variation of the maximum eigenstate expectation values $\Delta {\cal O}$, quantified over excited states, with respect to its ground-state average (see text) for different correlators, when taking into account at least 24 eigenstates in the low-lying spectrum for the ${\bf k} = (0,0)$ momentum sector. Parameters are the same as in Fig.~\ref{fig:static}: 
$U=8\,t_h$ and $n_e=14$.
}
\label{fig:relative_var} 
\end{figure}

Figure \ref{fig:static} displays the eigenstate expectation values (EEV) of the previously introduced correlators for $t_h'/t_h^{\phantom{'}}=0.8$, $U=8\,t_h$. Most noticeably, a group of three excited states (two of them degenerate) with energy gap $E_{\alpha}-E_0 \simeq 0.25\, t_h$  (belonging to the same momentum sector) has $d$-wave pairing $P_d \simeq 1.9$, significantly higher than that of the ground state $P_d=1.66$. With the exception of the staggered spin correlations, no similar substantial relative increase is seen in other correlators, including the extended $s$-wave or CDW channels. Equivalent simulations were performed for other $t_h^\prime/t_h^{\phantom{'}}$ values, and a compilation of the results of the relative maximum enhancement over excited states with respect to the ground-state, $\Delta {\cal O} \equiv (\max\{{\langle \alpha | \hat {\cal O} | \alpha\rangle\}}  - \langle 0|\hat  {\cal O} | 0\rangle)/\langle 0| \hat {\cal O} | 0\rangle$, is given in Fig.~\ref{fig:relative_var}. Although this maximum increase for the $d$-wave pairing  is seen right in the vicinity of the homogeneous system ($t^\prime_h=t_h$), we focus on $t_h'/t_h=0.8$ to study the regime of robustly formed plaquettes.

\section{Out-of-equilibrium results: Photoirradiation}
\begin{figure}[t!]
\includegraphics[width=1\columnwidth]{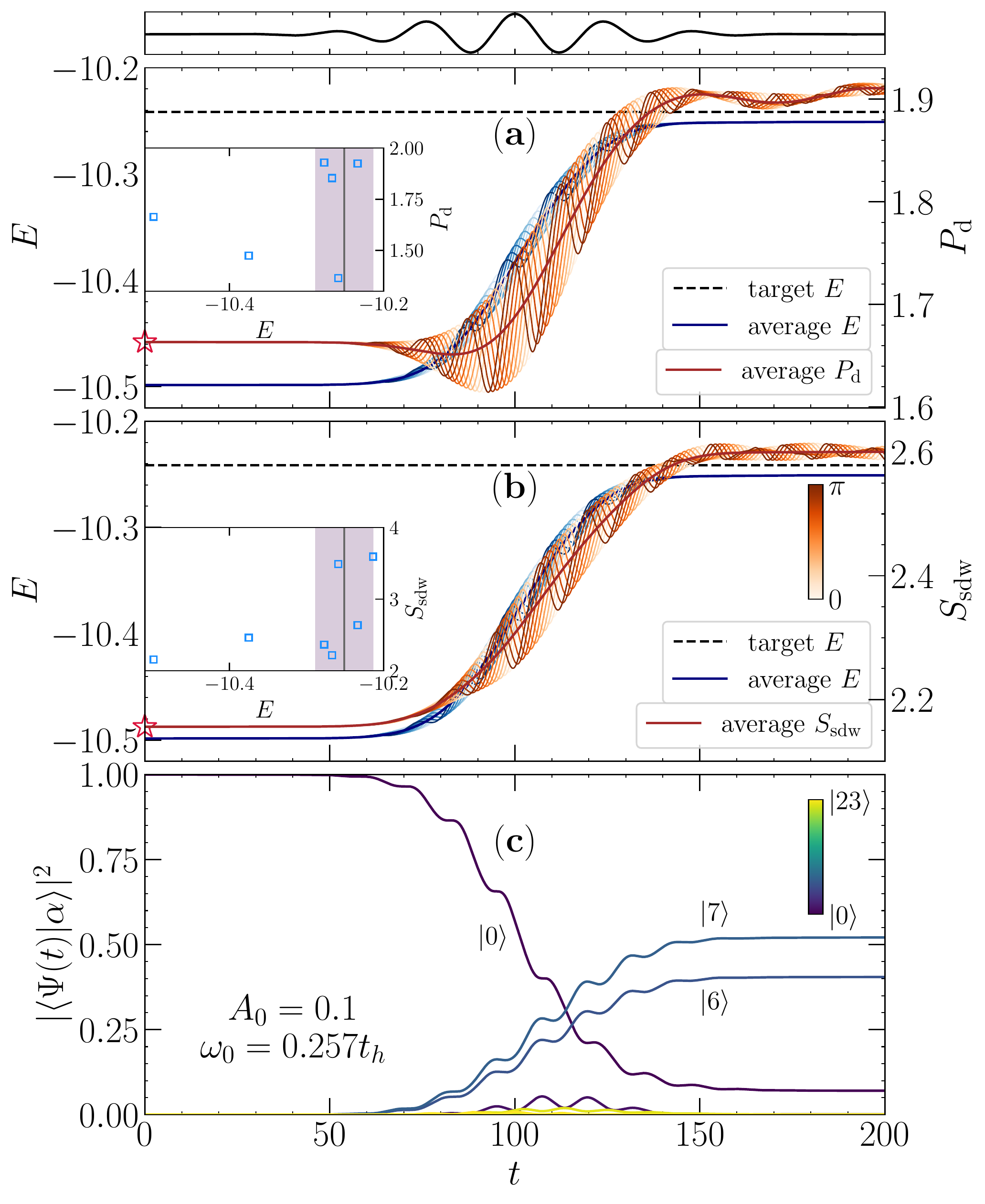} %
\caption{Top panel: schematics of the pulse amplitude over time. (a) Evolution of the phase-averaged $d$-wave pair correlation $P_d$ (red solid curve) and mean energy $E$ (blue solid curve). The different curves in shades of red and blue denote the values for each time-phase $\varphi$. (b) The same for the staggered spin structure factor. (c) Dynamics of the overlaps $|\langle \Psi(t)|\alpha \rangle |^2$ over the course of photoirradiation. Darker color curves represent lower energy states; the degenerate doublet, $|6\rangle$ and $7\rangle$, displays the largest contribution to $|\Psi(t)\rangle$ as marked, whereas the participation of $|0\rangle$, the initial state, diminishes significantly. The insets give the EEVs for both quantities with the mean energy at $t=t_{\rm max}$ and within a window given by $\sigma_{E}$ (see text). Pump parameters are $A_0=0.1$ and $\omega_0=0.257 \, t_h$. 
}
\label{fig:t_evol} 
\end{figure}

Having established that the expectation values of correlators, in particular $P_d$, measured in excited eigenstates of the Hamiltonian may possess larger values in comparison to the ones at the ground state, we turn to the out-of-equilibrium scenario, with an added time-dependent external field which can cause transitions to these states.
Its associated vector potential amplitude is depicted at the top of Fig.~\ref{fig:t_evol}. We follow a procedure that has been proven feasible in other contexts of photoirradiation~\cite{shao2019,shao21}: by using a fine-tuned temporal perturbation, whose frequency is set resonantly with the target state one is aiming to excite ($\omega_0 \equiv E_{\rm target} - E_0$), the dynamics at long times thus exhibit a large overlap with such eigenstate.

There are a number of constraints in this protocol. First, the initial and final states need similar quantum numbers. In particular, total momentum, as the vector potential used [Eq.~\eqref{eq:vec_pot}] is spatially uniform and does not break translation invariance. Second, the pump needs to be time-constrained, otherwise the system would indefinitely absorb energy from the drive, owing to the thermalizing properties of non-integrable systems~\cite{Alessio14}. In such case, heating towards an infinite temperature regime would ensue at sufficiently long-times, exhibiting featureless correlators.

Following these requirements, we use a pump with parameters $A_0=0.1$, $\omega_0=0.257 \, t_h$ as to closely match that of the excited states with stronger $d$-wave pairing signal [see top of Fig.~\ref{fig:static}(a)]. These parameters are also reasonably aligned to the ones employed in recent experiments. Using $t_h \simeq 0.3$ eV, which is most typically seen in most cuprates materials, $\omega_0=0.257 \, t_h$ corresponds to $0.08$ eV or 20 THz whereas $A_0=0.1$ translates into a pump fluence of $\sim 
0.02$ J/cm$^2$, which are comparable to that used in photo-induced experiments \cite{cavalleri2018,cremin2019}. A discussion of the neighboring frequencies that can influence the dynamics and the reason why we chose this specific $\omega_0$ is given in Appendix \ref{app:neighboring_freq}. One can, however, see from Fig.~\ref{fig:static} that this
$\omega_0$ at least roughly corresponds to the excitation energy of a set of states
with enhanced $P_d$.
In particular, a doublet of states ($|6\rangle$ and $|7\rangle$), energy-degenerate up to $10^{-12}\ t_h$, possesses one of the largest expectation values of the $d$-wave pairing $P_d$ ($\simeq 1.92$). Under these conditions, Fig.~\ref{fig:t_evol}(a) shows that a considerable enhancement in $d$-wave pairing is induced, accompanied by an energy increase, such that the mean energy at long times, $E = \langle \Psi(t=t_{\rm max})|\hat {\cal H}|\Psi(t=t_{\rm max})\rangle$, is close to the eigenenergy of the target states. Simultaneously, a similar large enhancement on the staggered spin correlations is also achieved [Fig.~\ref{fig:t_evol}(b)]. 

To demonstrate that the dynamics is being influenced by the target and its neighboring states at these time scales, we compute the overlaps of the time-evolving wave functions and the eigenstates of the equilibrium Hamiltonian, $|\langle \Psi(t)|\alpha \rangle |^2$. We report those in Fig.~\ref{fig:t_evol}(c), where one notices that the weight of the target doublet states rises up to 0.52 and 0.40 after the pump, for states $|7\rangle$ and $|6\rangle$, respectively, explaining thus the enhancement of $d$-wave pairing under these conditions. The participation of the ground state $|0\rangle$ decreases significantly, whereas the contribution of other eigenstates is mostly negligible. In the insets, the EEVs of $P_d$ and $S_{\rm sdw}$ are shown in this set of parameters, $(U/t_h, t^\prime_h/t_h^{\phantom{'}})= (8,0.8)$, together with the mean energy of the system $E(t_{\rm max})$.


A first account of which eigenstates can significantly contribute to the dynamics at a given time is given by the width in energy of $|\Psi(t)\rangle$, defined as $\sigma_E = [\langle \Psi(t)|H^2|\Psi(t) \rangle - \langle \Psi(t)|H|\Psi(t)\rangle ^2]^{1/2}$~\cite{Sorg14,Bauer15}. This is displayed as a shaded area surrounding $E(t_{\rm max})$ in the insets of Fig.~\ref{fig:t_evol}(a) and \ref{fig:t_evol}(b),
where we notice that the eigenstates within the window are the ones that may display a large overlap with $|\Psi(t)\rangle$ in Fig.~\ref{fig:t_evol}(c). While this window contains six eigenstates in the ${\bf k} =(0,0)$ quasi-momentum sector, not all of them affect the dynamics. Symmetry requirements related to how they couple to the current operator 
$\hat {\cal J}
=  \sum_{\langle \mathbf{i}, \mathbf{j}
  \rangle, \sigma} \big(-i t_{\mathbf{i},\mathbf{j}}^{\phantom{\dagger}}
\, \hat c^{\dagger}_{\mathbf{i} \sigma} \hat
c^{\phantom{\dagger}}_{\mathbf{j} \sigma} + {\rm H.c.} \big)$ dictate which ones will eventually couple to the external field. That is, besides energy requirements, overlaps $|\langle \alpha| \hat {\cal J} |0\rangle|^2$ also classify, in first order, the states most influencing the dynamics.  Even if the pulse maintains the quasi-momentum of the time-evolving wavefunction, the polarization employed is not sufficient to mix all different mirror symmetry sectors, and as a result some eigenstates must not participate in the dynamical evolution process.

\begin{figure}[hbpt]
\includegraphics[width=0.85\columnwidth]{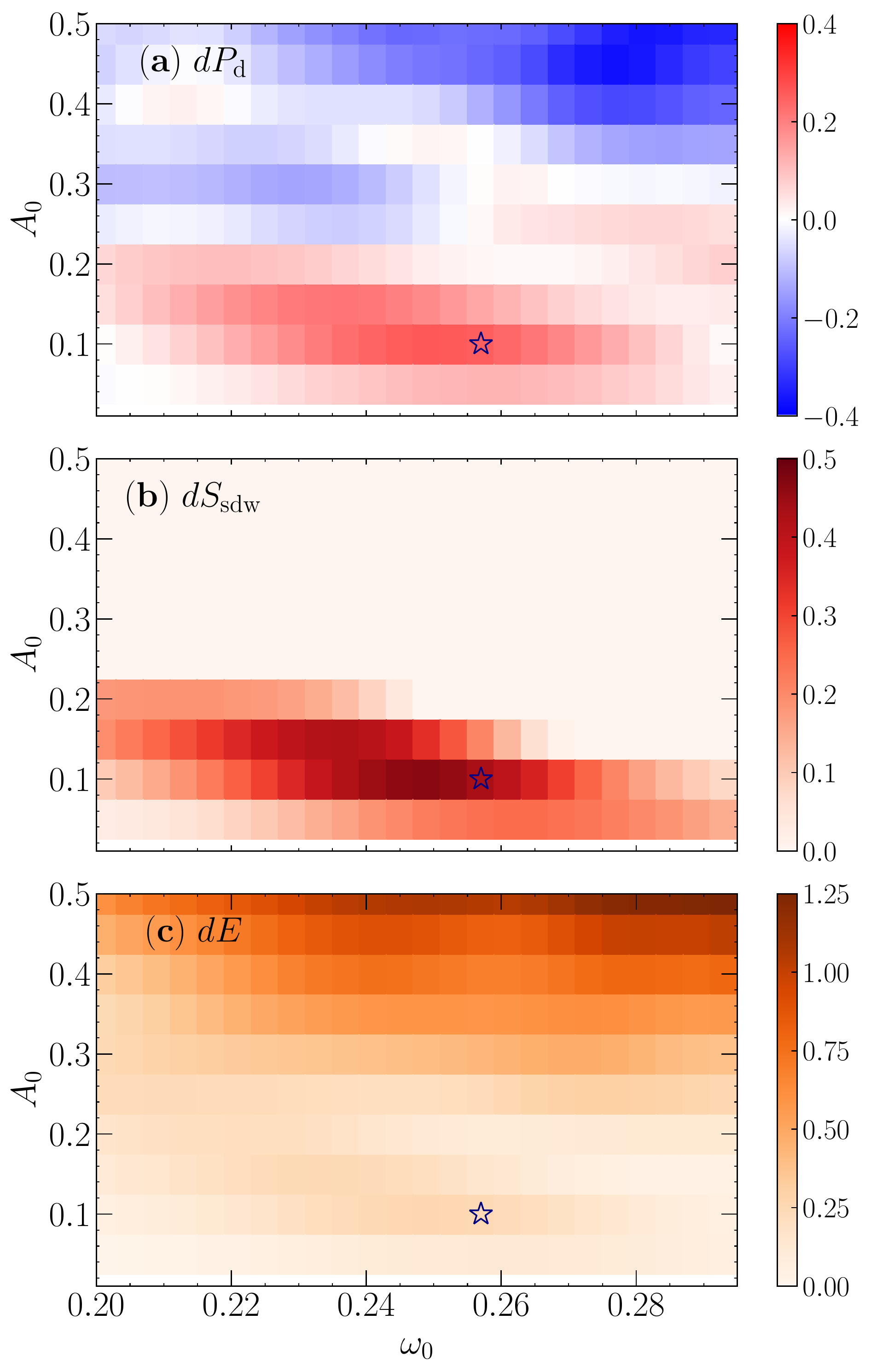} %
\caption{Contour plot of the enhancement of the $d$-wave pair [staggered spin] structure factor (a) [(b)] at $t=t_{\rm max}$ for various pump parameter $(A_0, \omega_0)$ values. (c) Similar results for the injected energy by the photoirradiation. All results are obtained for $\varphi = 0$. In all panels, star markers identify the pump parameters used in Fig.~\ref{fig:t_evol}.
}
\label{fig:colormap} 
\end{figure}

Now that we have confirmed that an enhancement of $d$-wave pairing is attainable 
for these pump parameters, we explore other combinations of $(A_0, \omega_0)$, identifying the ones that give optimal results in the enhancement of pair correlations at long times.  Towards that end, 
we study the variation of different post-pump observables as $dP_d \equiv P_d(t = t_{\rm max}) - P_d(0)$ [Fig.~\ref{fig:colormap}(a)] and $dS_{\rm sdw} \equiv S_{\rm sdw}(t = t_{\rm max}) - S_{\rm sdw}(0)$ [Fig.~\ref{fig:colormap}(b)], quantifying the dynamical variations of $d$-wave pairing and staggered spin correlations. A relatively narrow band in pump amplitudes $A_0$ results in increased pairing for resonant drives.  The reason  can be seen by the map of injected energies $dE \equiv E(t=t_{\rm max}) - E_0$ in Fig.~\ref{fig:colormap}(c): Increasing $A_0$ leads to a systematically larger absorbed energy by the system, thereby a larger contribution of states in the bulk of the spectrum, not associated to an enhanced pairing amplitude, is manifest.

We note that not only a specific resonant drive with frequency $\omega_0$, in principle, is able to target a state, but also higher harmonic contributions $2\, \omega_0, 3\, \omega_0, \ldots$.  A tuning of other parameters of the pulse, as $A_0$ and $t_d$, is necessary in these conditions in order to control the injected energy in the system, while monitoring the typical width in energy of $|\Psi(t_{\rm max})\rangle$. Adjusting these parameters, within physically reasonable limits, thus presents as a systematic protocol to access the physics of competing states close to the ground-state, guaranteeing that the long-time dynamics is largely influenced by the (group of) target excited state(s). Such entwining of different correlations is known to occur in doped Hubbard Hamiltonians and, as we demonstrated, engineered time-dependent perturbations are key to pick one (or more) type in detriment of others.


\section{Summary and Discussion}

We studied the plaquette Hubbard model under photoirradiation, with a goal of dynamical enhancement of $d$-wave superconductivity. By focusing on a regime away from half-filling ($\rho=0.875$), we observed a moderate increase of pairing (15\% for $t^\prime_h/t_h^{\phantom{'}}=0.8$) at long times in relation to the equilibrium ground state. This is accompanied by enhanced staggered spin fluctuations. Such correlated behavior is not unfamiliar to the physics of unconventional superconductors, where a close competition or entwining of states displaying spin and pairing order is known to occur. That we dynamically reach a simultaneous enhancement of both pair and spin degrees of freedom provides a further, interesting window into the interplay of the low-energy physics of the weakly doped Hubbard model.

A fundamental difference between our study and others that have numerically investigated the increase of superconducting correlations in time is that, in our case, the unperturbed Hamiltonian does not feature strong charge or spin fluctuations associated to long-range order. In particular, the unperturbed system is not a Mott insulator as in Ref.~\onlinecite{kaneko2019} nor a Peierls insulator as in Ref.~\onlinecite{wang2018}. Consequently, the gaps separating the ground state and others excited states are substantially smaller. This is one of the reasons behind our pump pulses having an approximate one order of magnitude smaller frequencies.

Our protocol, on the other hand, is intrinsically tied to the specific details of the Hamiltonian's low-lying spectrum, and can be used more generally to induce changes of the wave-function characteristics, surpassing the specific focus on superconductivity~\cite{shao2019,shao21}. It is worth noticing that a requirement to achieve an efficient targetting of states is that these are well separated in energy from the bulk of the spectrum, otherwise the overlaps with non-desired states is unavoidable. When approaching the thermodynamical limit (not addressed here) the low-lying eigenspectrum is still discrete (that is, does not form a continuous bulk), and identification of targeting protocol is still feasible. 

Nonetheless, a confirmation of these results in larger cluster sizes may shed light on the understanding of its impact in explaining potential connections to experimental results. We stress, however, that in dealing with small clusters, finite-size effects for $d$-wave correlators do not monotonic decrease with the system size, but are rather influenced by the number $z_d$ of independent neighboring $d$-wave plaquettes~\cite{Maier2005}. In the thermodynamic limit $z_d = 4$, and the smallest cluster size featuring such configuration possess $N=20$. This sits at the verge of what is computationally accessible within the scope of exact numerical methods. A potential venue of future investigation is to understand the competition of charge textures (as stripes) and pairing, now adding real-time dynamics to potentially enhance one degree of freedom in detriment of the other. This builds prospects to connect to recent experimental results in cuprates~\cite{fausti2011,nicoletti2014,Forst2014,cremin2019}.

A word of caution, however, is that realistic treatment of such materials using model Hamiltonians might necessarily take into account either phonon degrees of freedom~\cite{wang2021} or the effects of multiple bands~\cite{wang2021b},
such as the distinct Cu and O orbitals. Still, the `bottom-up' approach we develop to first understand the physics of the low-lying spectrum in order to justify dynamical modifications of suitable correlators should similarly work.
\appendix 

\section{The half-filled case} \label{app:half_fill}
In the main text, we focus on the case of a hole-doped system ($n_e =14$), guided by the enhanced superconducting properties in cuprate materials in these conditions, and by the large pair binding energy observed in a similar Hamiltonian~\cite{tsai2008}. Yet, the possibility that enhanced pairing can be obtained even when the parent system is a robust insulator has also been proven before~\cite{wang2018,kaneko2019}. Thus, using our guiding protocol to infer the likelihood that a photoirradiation perturbation, assessing first the properties of the low-lying spectrum,
we compute the largest relative enhancement of correlators at half-filling ($n_e=16$), in analogy to Fig.~\ref{fig:relative_var}.

\begin{figure}[hbpt]
\includegraphics[width=0.84\columnwidth]{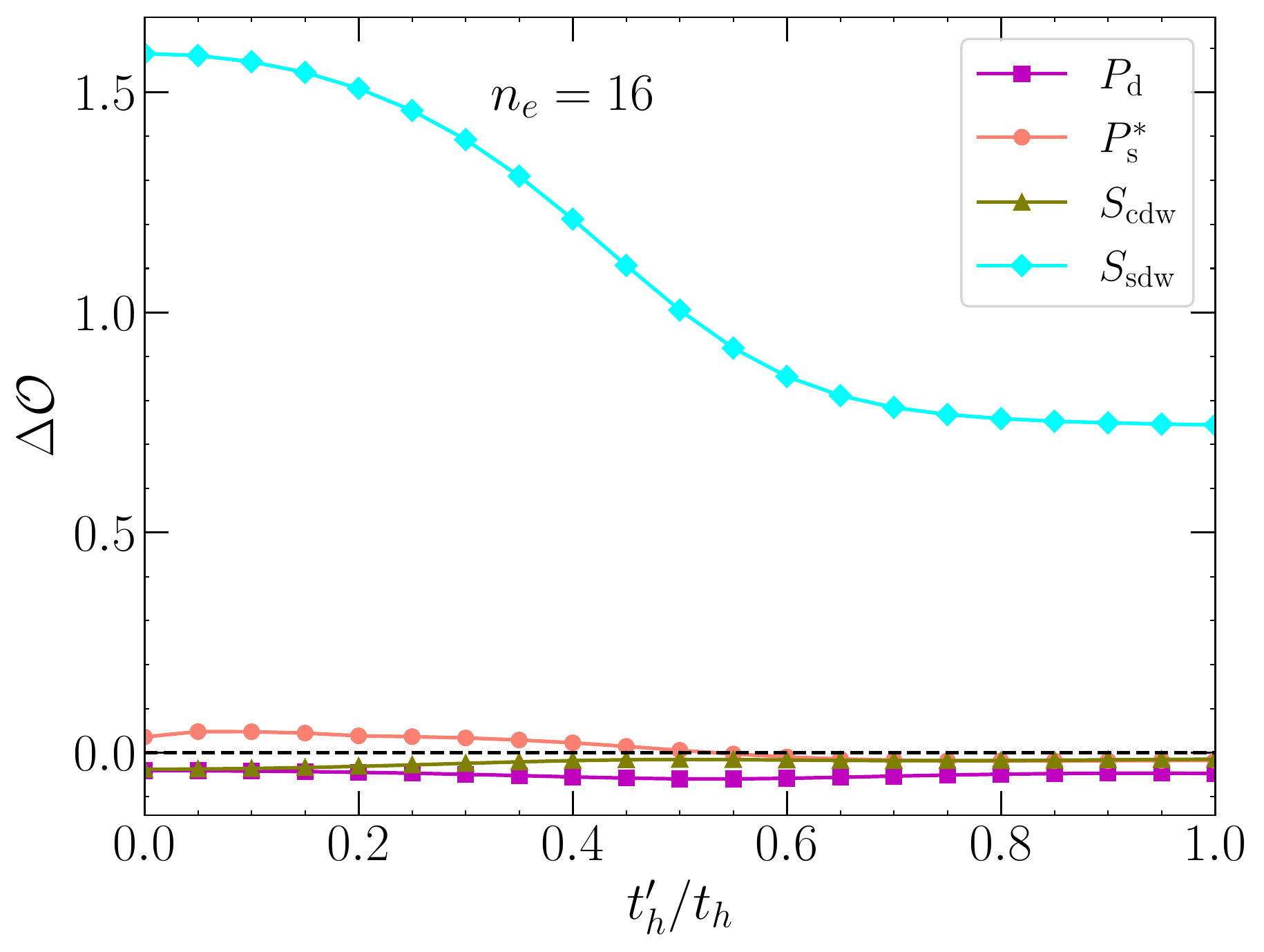} %
\caption{The equivalent of Fig.~\ref{fig:relative_var} but for the half-filled system at $U/t_h = 8$. Within a span of 24 eigenstates of the low-lying spectrum at zero quasi-momentum (the same where the ground-state resides) no siginificant increase of matrix elements for pairing correlators in respect to its ground-state average is observed. 
}
\label{fig:relative_increase_rho1} 
\end{figure}

We notice that other than potentially larger spin fluctuations, the remaining observables we investigate do not possess large matrix elements in the low-lying eigenspectrum in comparison to the ground-state. As a result, no significant dynamical enhancement of pairing is expected in this scenario,
including, specifically, $d$-wave pairing.

\section{Enhancement of $d$-wave pairing at neighboring frequencies} \label{app:neighboring_freq}
\begin{figure}[hbpt]
\includegraphics[width=0.84\columnwidth]{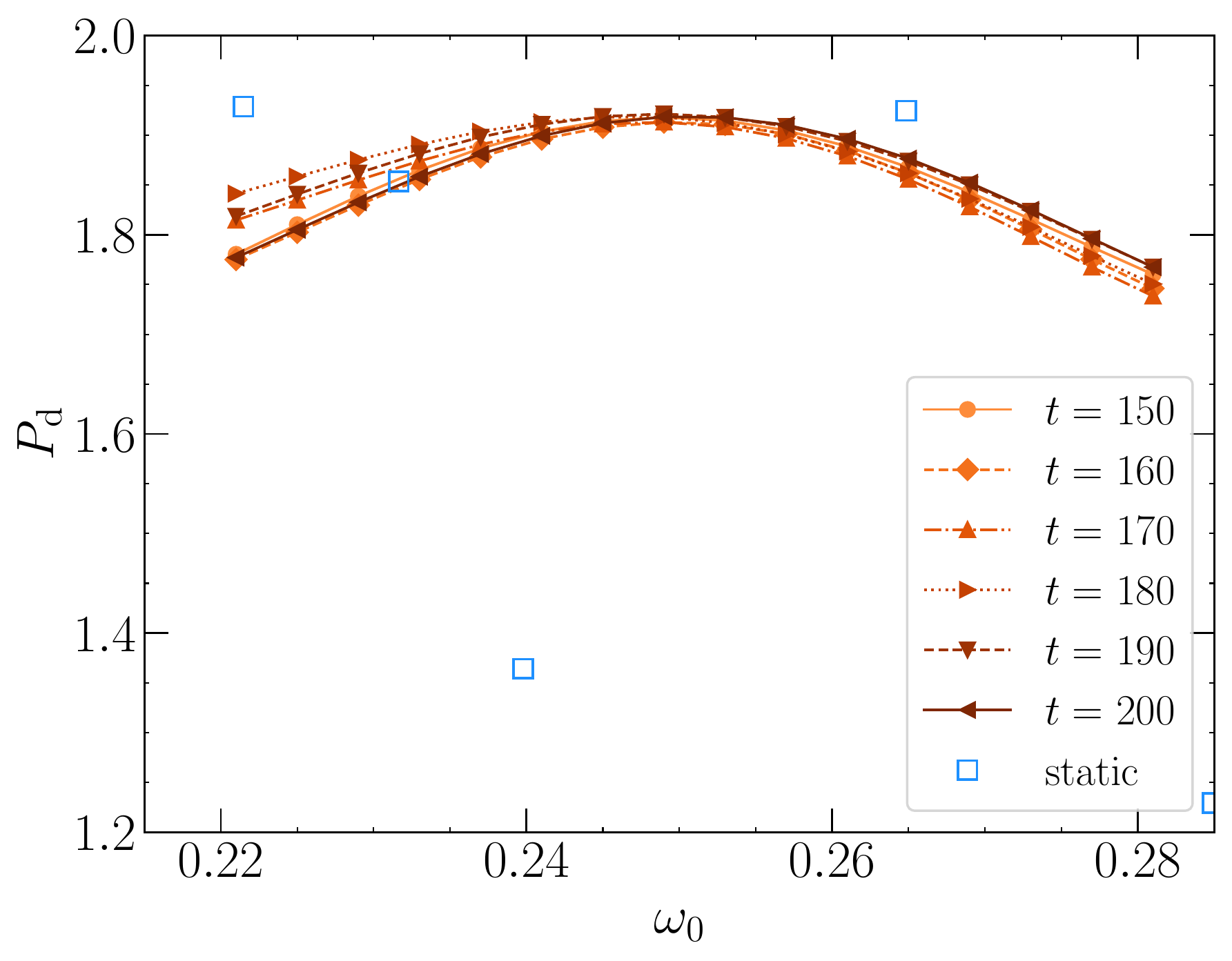} %
\caption{The phase averaged value of the $d$-wave pairing correlator at various late times as marked. Empty markers give the expectation values within various low-lying eigenstates. 
}
\label{fig:neighboring_frequencies} 
\end{figure}

In the main text we have used a specific pump frequency, $\omega_0=0.257 \, t_h$. Here we discuss the rationale behind this choice, based on the results for the phase averaged enhancement of the $P_d$ pairing correlators at different long times, as displayed in Fig.~\ref{fig:neighboring_frequencies}. The maximum enhancement of the $d$-wave pairing at times following the pulse are obtained for frequencies $\omega_0$ slightly below $E_{6,7}-E_0=0.2648 t_h$.  The reason is that, despite of the target doublet states being mostly excited, contributions from other states to the dynamics, as e.g.~state $|1\rangle$, lead to an optimal selection of pump parameters that are slightly below a simple locked-in frequency analysis at $E_{\rm target}-E_0$. 

In other contexts, when initial and target states are related by a first order phase transition in a slightly different regime of parameters of the Hamiltonian, resonant-selection of the states is often highly effective~\cite{shao2019,shao21}, since in such cases they are well separated from the bulk of the spectrum~\cite{Chen2010}. This may explain recent observations of enhancement of $d$-wave pairing in the vicinity of a phase boundary in systems where the interplay of electron and phonon degrees of freedom is present~\cite{wang2018,wang2021}.

\begin{acknowledgments}
\noindent
The work of YXZ, and RTS was supported by the grant DE‐SC0014671 funded by the U.S.~Department of Energy, Office of Science. R.M.~acknowledges support from the National Natural Science Foundation of China (NSFC) Grants No.~U1930402, 12050410263, 12111530010 and No.~11974039.  Computations were performed on the Tianhe-2JK at the Beijing  Computational  Science  Research  Center.
\end{acknowledgments}

\bibliography{tdepHub}

\begin{thebibliography}{45}%
\makeatletter
\providecommand \@ifxundefined [1]{%
 \@ifx{#1\undefined}
}%
\providecommand \@ifnum [1]{%
 \ifnum #1\expandafter \@firstoftwo
 \else \expandafter \@secondoftwo
 \fi
}%
\providecommand \@ifx [1]{%
 \ifx #1\expandafter \@firstoftwo
 \else \expandafter \@secondoftwo
 \fi
}%
\providecommand \natexlab [1]{#1}%
\providecommand \enquote  [1]{``#1''}%
\providecommand \bibnamefont  [1]{#1}%
\providecommand \bibfnamefont [1]{#1}%
\providecommand \citenamefont [1]{#1}%
\providecommand \href@noop [0]{\@secondoftwo}%
\providecommand \href [0]{\begingroup \@sanitize@url \@href}%
\providecommand \@href[1]{\@@startlink{#1}\@@href}%
\providecommand \@@href[1]{\endgroup#1\@@endlink}%
\providecommand \@sanitize@url [0]{\catcode `\\12\catcode `\$12\catcode
  `\&12\catcode `\#12\catcode `\^12\catcode `\_12\catcode `\%12\relax}%
\providecommand \@@startlink[1]{}%
\providecommand \@@endlink[0]{}%
\providecommand \url  [0]{\begingroup\@sanitize@url \@url }%
\providecommand \@url [1]{\endgroup\@href {#1}{\urlprefix }}%
\providecommand \urlprefix  [0]{URL }%
\providecommand \Eprint [0]{\href }%
\providecommand \doibase [0]{https://doi.org/}%
\providecommand \selectlanguage [0]{\@gobble}%
\providecommand \bibinfo  [0]{\@secondoftwo}%
\providecommand \bibfield  [0]{\@secondoftwo}%
\providecommand \translation [1]{[#1]}%
\providecommand \BibitemOpen [0]{}%
\providecommand \bibitemStop [0]{}%
\providecommand \bibitemNoStop [0]{.\EOS\space}%
\providecommand \EOS [0]{\spacefactor3000\relax}%
\providecommand \BibitemShut  [1]{\csname bibitem#1\endcsname}%
\let\auto@bib@innerbib\@empty
\bibitem [{\citenamefont {Oka}\ and\ \citenamefont {Kitamura}(2019)}]{oka19}%
  \BibitemOpen
  \bibfield  {author} {\bibinfo {author} {\bibfnamefont {T.}~\bibnamefont
  {Oka}}\ and\ \bibinfo {author} {\bibfnamefont {S.}~\bibnamefont {Kitamura}},\
  }\bibfield  {title} {\bibinfo {title} {Floquet engineering of quantum
  materials},\ }\href
  {https://doi.org/10.1146/annurev-conmatphys-031218-013423} {\bibfield
  {journal} {\bibinfo  {journal} {Annual Review of Condensed Matter Physics}\
  }\textbf {\bibinfo {volume} {10}},\ \bibinfo {pages} {387} (\bibinfo {year}
  {2019})}\BibitemShut {NoStop}%
\bibitem [{\citenamefont {Wang}\ \emph {et~al.}(2017)\citenamefont {Wang},
  \citenamefont {Claassen}, \citenamefont {Moritz},\ and\ \citenamefont
  {Devereaux}}]{wang2017}%
  \BibitemOpen
  \bibfield  {author} {\bibinfo {author} {\bibfnamefont {Y.}~\bibnamefont
  {Wang}}, \bibinfo {author} {\bibfnamefont {M.}~\bibnamefont {Claassen}},
  \bibinfo {author} {\bibfnamefont {B.}~\bibnamefont {Moritz}},\ and\ \bibinfo
  {author} {\bibfnamefont {T.}~\bibnamefont {Devereaux}},\ }\bibfield  {title}
  {\bibinfo {title} {Producing coherent excitations in pumped {M}ott
  antiferromagnetic insulators},\ }\href
  {https://doi.org/10.1103/PhysRevB.96.235142} {\bibfield  {journal} {\bibinfo
  {journal} {Physical Review B}\ }\textbf {\bibinfo {volume} {96}},\ \bibinfo
  {pages} {235142} (\bibinfo {year} {2017})}\BibitemShut {NoStop}%
\bibitem [{\citenamefont {Wang}\ \emph
  {et~al.}(2018{\natexlab{a}})\citenamefont {Wang}, \citenamefont {Claassen},
  \citenamefont {Pemmaraju}, \citenamefont {Jia}, \citenamefont {Moritz},\ and\
  \citenamefont {Devereaux}}]{wang2018t}%
  \BibitemOpen
  \bibfield  {author} {\bibinfo {author} {\bibfnamefont {Y.}~\bibnamefont
  {Wang}}, \bibinfo {author} {\bibfnamefont {M.}~\bibnamefont {Claassen}},
  \bibinfo {author} {\bibfnamefont {C.~D.}\ \bibnamefont {Pemmaraju}}, \bibinfo
  {author} {\bibfnamefont {C.}~\bibnamefont {Jia}}, \bibinfo {author}
  {\bibfnamefont {B.}~\bibnamefont {Moritz}},\ and\ \bibinfo {author}
  {\bibfnamefont {T.~P.}\ \bibnamefont {Devereaux}},\ }\bibfield  {title}
  {\bibinfo {title} {Theoretical understanding of photon spectroscopies in
  correlated materials in and out of equilibrium},\ }\href
  {https://doi.org/10.1038/s41578-018-0046-3} {\bibfield  {journal} {\bibinfo
  {journal} {Nature Reviews Materials}\ }\textbf {\bibinfo {volume} {3}},\
  \bibinfo {pages} {312} (\bibinfo {year} {2018}{\natexlab{a}})}\BibitemShut
  {NoStop}%
\bibitem [{\citenamefont {Kaneko}\ \emph {et~al.}(2019)\citenamefont {Kaneko},
  \citenamefont {Shirakawa}, \citenamefont {Sorella},\ and\ \citenamefont
  {Yunoki}}]{kaneko2019}%
  \BibitemOpen
  \bibfield  {author} {\bibinfo {author} {\bibfnamefont {T.}~\bibnamefont
  {Kaneko}}, \bibinfo {author} {\bibfnamefont {T.}~\bibnamefont {Shirakawa}},
  \bibinfo {author} {\bibfnamefont {S.}~\bibnamefont {Sorella}},\ and\ \bibinfo
  {author} {\bibfnamefont {S.}~\bibnamefont {Yunoki}},\ }\bibfield  {title}
  {\bibinfo {title} {Photoinduced $\eta$ pairing in the {H}ubbard model},\
  }\href {https://doi.org/10.1103/PhysRevLett.122.077002} {\bibfield  {journal}
  {\bibinfo  {journal} {Physical Review Letters}\ }\textbf {\bibinfo {volume}
  {122}},\ \bibinfo {pages} {077002} (\bibinfo {year} {2019})}\BibitemShut
  {NoStop}%
\bibitem [{\citenamefont {Shirakawa}\ \emph {et~al.}(2020)\citenamefont
  {Shirakawa}, \citenamefont {Miyakoshi},\ and\ \citenamefont
  {Yunoki}}]{kaneko2020}%
  \BibitemOpen
  \bibfield  {author} {\bibinfo {author} {\bibfnamefont {T.}~\bibnamefont
  {Shirakawa}}, \bibinfo {author} {\bibfnamefont {S.}~\bibnamefont
  {Miyakoshi}},\ and\ \bibinfo {author} {\bibfnamefont {S.}~\bibnamefont
  {Yunoki}},\ }\bibfield  {title} {\bibinfo {title} {Photoinduced
  $\ensuremath{\eta}$ pairing in the {K}ondo lattice model},\ }\href
  {https://doi.org/10.1103/PhysRevB.101.174307} {\bibfield  {journal} {\bibinfo
   {journal} {Phys. Rev. B}\ }\textbf {\bibinfo {volume} {101}},\ \bibinfo
  {pages} {174307} (\bibinfo {year} {2020})}\BibitemShut {NoStop}%
\bibitem [{\citenamefont {Shao}\ \emph {et~al.}(2019)\citenamefont {Shao},
  \citenamefont {Lu}, \citenamefont {Luo},\ and\ \citenamefont
  {Mondaini}}]{shao2019}%
  \BibitemOpen
  \bibfield  {author} {\bibinfo {author} {\bibfnamefont {C.}~\bibnamefont
  {Shao}}, \bibinfo {author} {\bibfnamefont {H.}~\bibnamefont {Lu}}, \bibinfo
  {author} {\bibfnamefont {H.-G.}\ \bibnamefont {Luo}},\ and\ \bibinfo {author}
  {\bibfnamefont {R.}~\bibnamefont {Mondaini}},\ }\bibfield  {title} {\bibinfo
  {title} {Photoinduced enhancement of bond order in the one-dimensional
  extended {H}ubbard model},\ }\href
  {https://doi.org/10.1103/PhysRevB.100.041114} {\bibfield  {journal} {\bibinfo
   {journal} {Physical Review B}\ }\textbf {\bibinfo {volume} {100}},\ \bibinfo
  {pages} {041114} (\bibinfo {year} {2019})}\BibitemShut {NoStop}%
\bibitem [{\citenamefont {McIver}\ \emph {et~al.}(2020)\citenamefont {McIver},
  \citenamefont {Schulte}, \citenamefont {Stein}, \citenamefont {Matsuyama},
  \citenamefont {Jotzu}, \citenamefont {Meier},\ and\ \citenamefont
  {Cavalleri}}]{mciver20}%
  \BibitemOpen
  \bibfield  {author} {\bibinfo {author} {\bibfnamefont {J.}~\bibnamefont
  {McIver}}, \bibinfo {author} {\bibfnamefont {B.}~\bibnamefont {Schulte}},
  \bibinfo {author} {\bibfnamefont {F.-U.}\ \bibnamefont {Stein}}, \bibinfo
  {author} {\bibfnamefont {T.}~\bibnamefont {Matsuyama}}, \bibinfo {author}
  {\bibfnamefont {G.}~\bibnamefont {Jotzu}}, \bibinfo {author} {\bibfnamefont
  {G.}~\bibnamefont {Meier}},\ and\ \bibinfo {author} {\bibfnamefont
  {A.}~\bibnamefont {Cavalleri}},\ }\bibfield  {title} {\bibinfo {title}
  {{Light-induced anomalous Hall effect in graphene}},\ }\href
  {https://doi.org/10.1038/s41567-019-0698-y} {\bibfield  {journal} {\bibinfo
  {journal} {Nature Physics}\ }\textbf {\bibinfo {volume} {16}},\ \bibinfo
  {pages} {38} (\bibinfo {year} {2020})}\BibitemShut {NoStop}%
\bibitem [{\citenamefont {Shao}\ \emph {et~al.}(2021)\citenamefont {Shao},
  \citenamefont {Sacramento},\ and\ \citenamefont {Mondaini}}]{shao21}%
  \BibitemOpen
  \bibfield  {author} {\bibinfo {author} {\bibfnamefont {C.}~\bibnamefont
  {Shao}}, \bibinfo {author} {\bibfnamefont {P.~D.}\ \bibnamefont
  {Sacramento}},\ and\ \bibinfo {author} {\bibfnamefont {R.}~\bibnamefont
  {Mondaini}},\ }\bibfield  {title} {\bibinfo {title} {Photoinduced anomalous
  {H}all effect in the interacting {H}aldane model: {T}argeting topological
  states with pump pulses},\ }\href
  {https://doi.org/10.1103/PhysRevB.104.125129} {\bibfield  {journal} {\bibinfo
   {journal} {Phys. Rev. B}\ }\textbf {\bibinfo {volume} {104}},\ \bibinfo
  {pages} {125129} (\bibinfo {year} {2021})}\BibitemShut {NoStop}%
\bibitem [{\citenamefont {Cavalleri}(2018)}]{cavalleri2018}%
  \BibitemOpen
  \bibfield  {author} {\bibinfo {author} {\bibfnamefont {A.}~\bibnamefont
  {Cavalleri}},\ }\bibfield  {title} {\bibinfo {title} {Photo-induced
  superconductivity},\ }\href {https://doi.org/10.1080/00107514.2017.1406623}
  {\bibfield  {journal} {\bibinfo  {journal} {Contemporary Physics}\ }\textbf
  {\bibinfo {volume} {59}},\ \bibinfo {pages} {31} (\bibinfo {year}
  {2018})}\BibitemShut {NoStop}%
\bibitem [{\citenamefont {Peronaci}\ \emph {et~al.}(2015)\citenamefont
  {Peronaci}, \citenamefont {Schir{\'o}},\ and\ \citenamefont
  {Capone}}]{peronaci2015}%
  \BibitemOpen
  \bibfield  {author} {\bibinfo {author} {\bibfnamefont {F.}~\bibnamefont
  {Peronaci}}, \bibinfo {author} {\bibfnamefont {M.}~\bibnamefont
  {Schir{\'o}}},\ and\ \bibinfo {author} {\bibfnamefont {M.}~\bibnamefont
  {Capone}},\ }\bibfield  {title} {\bibinfo {title} {Transient dynamics of
  $d$-wave superconductors after a sudden excitation},\ }\href
  {https://doi.org/10.1103/PhysRevLett.115.257001} {\bibfield  {journal}
  {\bibinfo  {journal} {Physical Review Letters}\ }\textbf {\bibinfo {volume}
  {115}},\ \bibinfo {pages} {257001} (\bibinfo {year} {2015})}\BibitemShut
  {NoStop}%
\bibitem [{\citenamefont {Wang}\ \emph {et~al.}(2016)\citenamefont {Wang},
  \citenamefont {Moritz}, \citenamefont {Chen}, \citenamefont {Jia},
  \citenamefont {van Veenendaal},\ and\ \citenamefont {Devereaux}}]{wang2016}%
  \BibitemOpen
  \bibfield  {author} {\bibinfo {author} {\bibfnamefont {Y.}~\bibnamefont
  {Wang}}, \bibinfo {author} {\bibfnamefont {B.}~\bibnamefont {Moritz}},
  \bibinfo {author} {\bibfnamefont {C.-C.}\ \bibnamefont {Chen}}, \bibinfo
  {author} {\bibfnamefont {C.}~\bibnamefont {Jia}}, \bibinfo {author}
  {\bibfnamefont {M.}~\bibnamefont {van Veenendaal}},\ and\ \bibinfo {author}
  {\bibfnamefont {T.~P.}\ \bibnamefont {Devereaux}},\ }\bibfield  {title}
  {\bibinfo {title} {Using nonequilibrium dynamics to probe competing orders in
  a {M}ott-{P}eierls system},\ }\href
  {https://doi.org/10.1103/PhysRevLett.116.086401} {\bibfield  {journal}
  {\bibinfo  {journal} {Physical Review Letters}\ }\textbf {\bibinfo {volume}
  {116}},\ \bibinfo {pages} {086401} (\bibinfo {year} {2016})}\BibitemShut
  {NoStop}%
\bibitem [{\citenamefont {Sentef}\ \emph {et~al.}(2017)\citenamefont {Sentef},
  \citenamefont {Tokuno}, \citenamefont {Georges},\ and\ \citenamefont
  {Kollath}}]{sentef2017}%
  \BibitemOpen
  \bibfield  {author} {\bibinfo {author} {\bibfnamefont {M.~A.}\ \bibnamefont
  {Sentef}}, \bibinfo {author} {\bibfnamefont {A.}~\bibnamefont {Tokuno}},
  \bibinfo {author} {\bibfnamefont {A.}~\bibnamefont {Georges}},\ and\ \bibinfo
  {author} {\bibfnamefont {C.}~\bibnamefont {Kollath}},\ }\bibfield  {title}
  {\bibinfo {title} {Theory of laser-controlled competing superconducting and
  charge orders},\ }\href {https://doi.org/10.1103/PhysRevLett.118.087002}
  {\bibfield  {journal} {\bibinfo  {journal} {Phys. Rev. Lett.}\ }\textbf
  {\bibinfo {volume} {118}},\ \bibinfo {pages} {087002} (\bibinfo {year}
  {2017})}\BibitemShut {NoStop}%
\bibitem [{\citenamefont {Wang}\ \emph
  {et~al.}(2018{\natexlab{b}})\citenamefont {Wang}, \citenamefont {Chen},
  \citenamefont {Moritz},\ and\ \citenamefont {Devereaux}}]{wang2018}%
  \BibitemOpen
  \bibfield  {author} {\bibinfo {author} {\bibfnamefont {Y.}~\bibnamefont
  {Wang}}, \bibinfo {author} {\bibfnamefont {C.-C.}\ \bibnamefont {Chen}},
  \bibinfo {author} {\bibfnamefont {B.}~\bibnamefont {Moritz}},\ and\ \bibinfo
  {author} {\bibfnamefont {T.}~\bibnamefont {Devereaux}},\ }\bibfield  {title}
  {\bibinfo {title} {Light-enhanced spin fluctuations and d-wave
  superconductivity at a phase boundary},\ }\href
  {https://doi.org/10.1103/PhysRevLett.120.246402} {\bibfield  {journal}
  {\bibinfo  {journal} {Physical Review Letters}\ }\textbf {\bibinfo {volume}
  {120}},\ \bibinfo {pages} {246402} (\bibinfo {year}
  {2018}{\natexlab{b}})}\BibitemShut {NoStop}%
\bibitem [{\citenamefont {Wang}\ \emph
  {et~al.}(2021{\natexlab{a}})\citenamefont {Wang}, \citenamefont {Shi},\ and\
  \citenamefont {Chen}}]{wang2021}%
  \BibitemOpen
  \bibfield  {author} {\bibinfo {author} {\bibfnamefont {Y.}~\bibnamefont
  {Wang}}, \bibinfo {author} {\bibfnamefont {T.}~\bibnamefont {Shi}},\ and\
  \bibinfo {author} {\bibfnamefont {C.-C.}\ \bibnamefont {Chen}},\ }\bibfield
  {title} {\bibinfo {title} {Fluctuating nature of light-enhanced $d$-wave
  superconductivity: A time-dependent variational non-{G}aussian exact
  diagonalization study},\ }\href {https://doi.org/10.1103/PhysRevX.11.041028}
  {\bibfield  {journal} {\bibinfo  {journal} {Phys. Rev. X}\ }\textbf {\bibinfo
  {volume} {11}},\ \bibinfo {pages} {041028} (\bibinfo {year}
  {2021}{\natexlab{a}})}\BibitemShut {NoStop}%
\bibitem [{\citenamefont {Tang}\ \emph {et~al.}(2021)\citenamefont {Tang},
  \citenamefont {Wang}, \citenamefont {Moritz},\ and\ \citenamefont
  {Devereaux}}]{tang2021}%
  \BibitemOpen
  \bibfield  {author} {\bibinfo {author} {\bibfnamefont {T.}~\bibnamefont
  {Tang}}, \bibinfo {author} {\bibfnamefont {Y.}~\bibnamefont {Wang}}, \bibinfo
  {author} {\bibfnamefont {B.}~\bibnamefont {Moritz}},\ and\ \bibinfo {author}
  {\bibfnamefont {T.~P.}\ \bibnamefont {Devereaux}},\ }\bibfield  {title}
  {\bibinfo {title} {Orbitally selective resonant photodoping to enhance
  superconductivity},\ }\href {https://arxiv.org/abs/2106.04742} {\bibfield
  {journal} {\bibinfo  {journal} {arXiv preprint arXiv:2106.04742}\ } (\bibinfo
  {year} {2021})}\BibitemShut {NoStop}%
\bibitem [{\citenamefont {Fausti}\ \emph {et~al.}(2011)\citenamefont {Fausti},
  \citenamefont {Tobey}, \citenamefont {Dean}, \citenamefont {Kaiser},
  \citenamefont {Dienst}, \citenamefont {Hoffmann}, \citenamefont {Pyon},
  \citenamefont {Takayama}, \citenamefont {Takagi},\ and\ \citenamefont
  {Cavalleri}}]{fausti2011}%
  \BibitemOpen
  \bibfield  {author} {\bibinfo {author} {\bibfnamefont {D.}~\bibnamefont
  {Fausti}}, \bibinfo {author} {\bibfnamefont {R.}~\bibnamefont {Tobey}},
  \bibinfo {author} {\bibfnamefont {N.}~\bibnamefont {Dean}}, \bibinfo {author}
  {\bibfnamefont {S.}~\bibnamefont {Kaiser}}, \bibinfo {author} {\bibfnamefont
  {A.}~\bibnamefont {Dienst}}, \bibinfo {author} {\bibfnamefont {M.~C.}\
  \bibnamefont {Hoffmann}}, \bibinfo {author} {\bibfnamefont {S.}~\bibnamefont
  {Pyon}}, \bibinfo {author} {\bibfnamefont {T.}~\bibnamefont {Takayama}},
  \bibinfo {author} {\bibfnamefont {H.}~\bibnamefont {Takagi}},\ and\ \bibinfo
  {author} {\bibfnamefont {A.}~\bibnamefont {Cavalleri}},\ }\bibfield  {title}
  {\bibinfo {title} {Light-induced superconductivity in a stripe-ordered
  cuprate},\ }\href {https://doi.org/10.1126/science.1197294} {\bibfield
  {journal} {\bibinfo  {journal} {Science}\ }\textbf {\bibinfo {volume}
  {331}},\ \bibinfo {pages} {189} (\bibinfo {year} {2011})}\BibitemShut
  {NoStop}%
\bibitem [{\citenamefont {Kaiser}\ \emph {et~al.}(2014)\citenamefont {Kaiser},
  \citenamefont {Hunt}, \citenamefont {Nicoletti}, \citenamefont {Hu},
  \citenamefont {Gierz}, \citenamefont {Liu}, \citenamefont {Le~Tacon},
  \citenamefont {Loew}, \citenamefont {Haug}, \citenamefont {Keimer},\ and\
  \citenamefont {Cavalleri}}]{kaiser2014}%
  \BibitemOpen
  \bibfield  {author} {\bibinfo {author} {\bibfnamefont {S.}~\bibnamefont
  {Kaiser}}, \bibinfo {author} {\bibfnamefont {C.~R.}\ \bibnamefont {Hunt}},
  \bibinfo {author} {\bibfnamefont {D.}~\bibnamefont {Nicoletti}}, \bibinfo
  {author} {\bibfnamefont {W.}~\bibnamefont {Hu}}, \bibinfo {author}
  {\bibfnamefont {I.}~\bibnamefont {Gierz}}, \bibinfo {author} {\bibfnamefont
  {H.~Y.}\ \bibnamefont {Liu}}, \bibinfo {author} {\bibfnamefont
  {M.}~\bibnamefont {Le~Tacon}}, \bibinfo {author} {\bibfnamefont
  {T.}~\bibnamefont {Loew}}, \bibinfo {author} {\bibfnamefont {D.}~\bibnamefont
  {Haug}}, \bibinfo {author} {\bibfnamefont {B.}~\bibnamefont {Keimer}},\ and\
  \bibinfo {author} {\bibfnamefont {A.}~\bibnamefont {Cavalleri}},\ }\bibfield
  {title} {\bibinfo {title} {Optically induced coherent transport far above
  ${T}_{c}$ in underdoped {YB}a$_{2}${C}$_{3}${O}$_{6+\delta}$},\ }\href
  {https://doi.org/10.1103/PhysRevB.89.184516} {\bibfield  {journal} {\bibinfo
  {journal} {Phys. Rev. B}\ }\textbf {\bibinfo {volume} {89}},\ \bibinfo
  {pages} {184516} (\bibinfo {year} {2014})}\BibitemShut {NoStop}%
\bibitem [{\citenamefont {Nicoletti}\ \emph {et~al.}(2014)\citenamefont
  {Nicoletti}, \citenamefont {Casandruc}, \citenamefont {Laplace},
  \citenamefont {Khanna}, \citenamefont {Hunt}, \citenamefont {Kaiser},
  \citenamefont {Dhesi}, \citenamefont {Gu}, \citenamefont {Hill},\ and\
  \citenamefont {Cavalleri}}]{nicoletti2014}%
  \BibitemOpen
  \bibfield  {author} {\bibinfo {author} {\bibfnamefont {D.}~\bibnamefont
  {Nicoletti}}, \bibinfo {author} {\bibfnamefont {E.}~\bibnamefont
  {Casandruc}}, \bibinfo {author} {\bibfnamefont {Y.}~\bibnamefont {Laplace}},
  \bibinfo {author} {\bibfnamefont {V.}~\bibnamefont {Khanna}}, \bibinfo
  {author} {\bibfnamefont {C.~R.}\ \bibnamefont {Hunt}}, \bibinfo {author}
  {\bibfnamefont {S.}~\bibnamefont {Kaiser}}, \bibinfo {author} {\bibfnamefont
  {S.}~\bibnamefont {Dhesi}}, \bibinfo {author} {\bibfnamefont
  {G.}~\bibnamefont {Gu}}, \bibinfo {author} {\bibfnamefont {J.}~\bibnamefont
  {Hill}},\ and\ \bibinfo {author} {\bibfnamefont {A.}~\bibnamefont
  {Cavalleri}},\ }\bibfield  {title} {\bibinfo {title} {Optically induced
  superconductivity in striped {L}a$_{2-x}${B}a$_x${C}u{O}$_4$ by
  polarization-selective excitation in the near infrared},\ }\href
  {https://doi.org/10.1103/PhysRevB.90.100503} {\bibfield  {journal} {\bibinfo
  {journal} {Physical Review B}\ }\textbf {\bibinfo {volume} {90}},\ \bibinfo
  {pages} {100503} (\bibinfo {year} {2014})}\BibitemShut {NoStop}%
\bibitem [{\citenamefont {Cremin}\ \emph {et~al.}(2019)\citenamefont {Cremin},
  \citenamefont {Zhang}, \citenamefont {Homes}, \citenamefont {Gu},
  \citenamefont {Sun}, \citenamefont {Fogler}, \citenamefont {Millis},
  \citenamefont {Basov},\ and\ \citenamefont {Averitt}}]{cremin2019}%
  \BibitemOpen
  \bibfield  {author} {\bibinfo {author} {\bibfnamefont {K.~A.}\ \bibnamefont
  {Cremin}}, \bibinfo {author} {\bibfnamefont {J.}~\bibnamefont {Zhang}},
  \bibinfo {author} {\bibfnamefont {C.~C.}\ \bibnamefont {Homes}}, \bibinfo
  {author} {\bibfnamefont {G.~D.}\ \bibnamefont {Gu}}, \bibinfo {author}
  {\bibfnamefont {Z.}~\bibnamefont {Sun}}, \bibinfo {author} {\bibfnamefont
  {M.~M.}\ \bibnamefont {Fogler}}, \bibinfo {author} {\bibfnamefont {A.~J.}\
  \bibnamefont {Millis}}, \bibinfo {author} {\bibfnamefont {D.~N.}\
  \bibnamefont {Basov}},\ and\ \bibinfo {author} {\bibfnamefont {R.~D.}\
  \bibnamefont {Averitt}},\ }\bibfield  {title} {\bibinfo {title}
  {Photoenhanced metastable c-axis electrodynamics in stripe-ordered cuprate
  {L}a$_{1.885}${B}a$_{0.115}${C}u{O}$_{4}$},\ }\href
  {https://doi.org/10.1073/pnas.1908368116} {\bibfield  {journal} {\bibinfo
  {journal} {Proceedings of the National Academy of Sciences}\ }\textbf
  {\bibinfo {volume} {116}},\ \bibinfo {pages} {19875} (\bibinfo {year}
  {2019})}\BibitemShut {NoStop}%
\bibitem [{\citenamefont {Tsai}\ and\ \citenamefont
  {Kivelson}(2006)}]{tsai2006}%
  \BibitemOpen
  \bibfield  {author} {\bibinfo {author} {\bibfnamefont {W.-F.}\ \bibnamefont
  {Tsai}}\ and\ \bibinfo {author} {\bibfnamefont {S.~A.}\ \bibnamefont
  {Kivelson}},\ }\bibfield  {title} {\bibinfo {title} {Superconductivity in
  inhomogeneous {H}ubbard models},\ }\href
  {https://doi.org/10.1103/PhysRevB.73.214510} {\bibfield  {journal} {\bibinfo
  {journal} {Physical Review B}\ }\textbf {\bibinfo {volume} {73}},\ \bibinfo
  {pages} {214510} (\bibinfo {year} {2006})}\BibitemShut {NoStop}%
\bibitem [{\citenamefont {Yao}\ \emph {et~al.}(2007)\citenamefont {Yao},
  \citenamefont {Tsai},\ and\ \citenamefont {Kivelson}}]{yao2007}%
  \BibitemOpen
  \bibfield  {author} {\bibinfo {author} {\bibfnamefont {H.}~\bibnamefont
  {Yao}}, \bibinfo {author} {\bibfnamefont {W.-F.}\ \bibnamefont {Tsai}},\ and\
  \bibinfo {author} {\bibfnamefont {S.~A.}\ \bibnamefont {Kivelson}},\
  }\bibfield  {title} {\bibinfo {title} {Myriad phases of the checkerboard
  {H}ubbard model},\ }\href {https://doi.org/10.1103/PhysRevB.76.161104}
  {\bibfield  {journal} {\bibinfo  {journal} {Physical Review B}\ }\textbf
  {\bibinfo {volume} {76}},\ \bibinfo {pages} {161104} (\bibinfo {year}
  {2007})}\BibitemShut {NoStop}%
\bibitem [{\citenamefont {Tsai}\ \emph {et~al.}(2008)\citenamefont {Tsai},
  \citenamefont {Yao}, \citenamefont {L{\"a}uchli},\ and\ \citenamefont
  {Kivelson}}]{tsai2008}%
  \BibitemOpen
  \bibfield  {author} {\bibinfo {author} {\bibfnamefont {W.-F.}\ \bibnamefont
  {Tsai}}, \bibinfo {author} {\bibfnamefont {H.}~\bibnamefont {Yao}}, \bibinfo
  {author} {\bibfnamefont {A.}~\bibnamefont {L{\"a}uchli}},\ and\ \bibinfo
  {author} {\bibfnamefont {S.~A.}\ \bibnamefont {Kivelson}},\ }\bibfield
  {title} {\bibinfo {title} {Optimal inhomogeneity for superconductivity:
  Finite-size studies},\ }\href {https://doi.org/10.1103/PhysRevB.77.214502}
  {\bibfield  {journal} {\bibinfo  {journal} {Physical Review B}\ }\textbf
  {\bibinfo {volume} {77}},\ \bibinfo {pages} {214502} (\bibinfo {year}
  {2008})}\BibitemShut {NoStop}%
\bibitem [{\citenamefont {Doluweera}\ \emph {et~al.}(2008)\citenamefont
  {Doluweera}, \citenamefont {Macridin}, \citenamefont {Maier}, \citenamefont
  {Jarrell},\ and\ \citenamefont {Pruschke}}]{doluweera2008}%
  \BibitemOpen
  \bibfield  {author} {\bibinfo {author} {\bibfnamefont {D.}~\bibnamefont
  {Doluweera}}, \bibinfo {author} {\bibfnamefont {A.}~\bibnamefont {Macridin}},
  \bibinfo {author} {\bibfnamefont {T.}~\bibnamefont {Maier}}, \bibinfo
  {author} {\bibfnamefont {M.}~\bibnamefont {Jarrell}},\ and\ \bibinfo {author}
  {\bibfnamefont {T.}~\bibnamefont {Pruschke}},\ }\bibfield  {title} {\bibinfo
  {title} {Suppression of $d$-wave superconductivity in the checkerboard
  {H}ubbard model},\ }\href {https://doi.org/10.1103/PhysRevB.78.020504}
  {\bibfield  {journal} {\bibinfo  {journal} {Physical Review B}\ }\textbf
  {\bibinfo {volume} {78}},\ \bibinfo {pages} {020504} (\bibinfo {year}
  {2008})}\BibitemShut {NoStop}%
\bibitem [{\citenamefont {Yao}\ and\ \citenamefont {Kivelson}(2010)}]{yao2010}%
  \BibitemOpen
  \bibfield  {author} {\bibinfo {author} {\bibfnamefont {H.}~\bibnamefont
  {Yao}}\ and\ \bibinfo {author} {\bibfnamefont {S.~A.}\ \bibnamefont
  {Kivelson}},\ }\bibfield  {title} {\bibinfo {title} {Fragile {M}ott
  insulators},\ }\href {https://doi.org/10.1103/PhysRevLett.105.166402}
  {\bibfield  {journal} {\bibinfo  {journal} {Physical Review Letters}\
  }\textbf {\bibinfo {volume} {105}},\ \bibinfo {pages} {166402} (\bibinfo
  {year} {2010})}\BibitemShut {NoStop}%
\bibitem [{\citenamefont {Smith}\ and\ \citenamefont
  {Kennett}(2013)}]{smith13}%
  \BibitemOpen
  \bibfield  {author} {\bibinfo {author} {\bibfnamefont {P.~M.}\ \bibnamefont
  {Smith}}\ and\ \bibinfo {author} {\bibfnamefont {M.~P.}\ \bibnamefont
  {Kennett}},\ }\bibfield  {title} {\bibinfo {title} {Disorder effects on
  superconducting tendencies in the checkerboard hubbard model},\ }\href
  {https://doi.org/10.1103/PhysRevB.88.214518} {\bibfield  {journal} {\bibinfo
  {journal} {Phys. Rev. B}\ }\textbf {\bibinfo {volume} {88}},\ \bibinfo
  {pages} {214518} (\bibinfo {year} {2013})}\BibitemShut {NoStop}%
\bibitem [{\citenamefont {Ying}\ \emph {et~al.}(2014)\citenamefont {Ying},
  \citenamefont {Mondaini}, \citenamefont {Sun}, \citenamefont {Paiva},
  \citenamefont {Fye},\ and\ \citenamefont {Scalettar}}]{ying2014}%
  \BibitemOpen
  \bibfield  {author} {\bibinfo {author} {\bibfnamefont {T.}~\bibnamefont
  {Ying}}, \bibinfo {author} {\bibfnamefont {R.}~\bibnamefont {Mondaini}},
  \bibinfo {author} {\bibfnamefont {X.}~\bibnamefont {Sun}}, \bibinfo {author}
  {\bibfnamefont {T.}~\bibnamefont {Paiva}}, \bibinfo {author} {\bibfnamefont
  {R.}~\bibnamefont {Fye}},\ and\ \bibinfo {author} {\bibfnamefont
  {R.}~\bibnamefont {Scalettar}},\ }\bibfield  {title} {\bibinfo {title}
  {Determinant quantum {M}onte {C}arlo study of $d$-wave pairing in the
  plaquette {H}ubbard {H}amiltonian},\ }\href
  {https://doi.org/10.1103/PhysRevB.90.075121} {\bibfield  {journal} {\bibinfo
  {journal} {Physical Review B}\ }\textbf {\bibinfo {volume} {90}},\ \bibinfo
  {pages} {075121} (\bibinfo {year} {2014})}\BibitemShut {NoStop}%
\bibitem [{\citenamefont {Kivelson}\ \emph {et~al.}(2003)\citenamefont
  {Kivelson}, \citenamefont {Bindloss}, \citenamefont {Fradkin}, \citenamefont
  {Oganesyan}, \citenamefont {Tranquada}, \citenamefont {Kapitulnik},\ and\
  \citenamefont {Howald}}]{kivelson03}%
  \BibitemOpen
  \bibfield  {author} {\bibinfo {author} {\bibfnamefont {S.~A.}\ \bibnamefont
  {Kivelson}}, \bibinfo {author} {\bibfnamefont {I.~P.}\ \bibnamefont
  {Bindloss}}, \bibinfo {author} {\bibfnamefont {E.}~\bibnamefont {Fradkin}},
  \bibinfo {author} {\bibfnamefont {V.}~\bibnamefont {Oganesyan}}, \bibinfo
  {author} {\bibfnamefont {J.}~\bibnamefont {Tranquada}}, \bibinfo {author}
  {\bibfnamefont {A.}~\bibnamefont {Kapitulnik}},\ and\ \bibinfo {author}
  {\bibfnamefont {C.}~\bibnamefont {Howald}},\ }\bibfield  {title} {\bibinfo
  {title} {How to detect fluctuating stripes in the high-temperature
  superconductors},\ }\href {https://doi.org/10.1103/RevModPhys.75.1201}
  {\bibfield  {journal} {\bibinfo  {journal} {Reviews of Modern Physics}\
  }\textbf {\bibinfo {volume} {75}},\ \bibinfo {pages} {1201} (\bibinfo {year}
  {2003})}\BibitemShut {NoStop}%
\bibitem [{\citenamefont {Berg}\ \emph {et~al.}(2009)\citenamefont {Berg},
  \citenamefont {Fradkin}, \citenamefont {Kivelson},\ and\ \citenamefont
  {Tranquada}}]{berg09}%
  \BibitemOpen
  \bibfield  {author} {\bibinfo {author} {\bibfnamefont {E.}~\bibnamefont
  {Berg}}, \bibinfo {author} {\bibfnamefont {E.}~\bibnamefont {Fradkin}},
  \bibinfo {author} {\bibfnamefont {S.~A.}\ \bibnamefont {Kivelson}},\ and\
  \bibinfo {author} {\bibfnamefont {J.~M.}\ \bibnamefont {Tranquada}},\
  }\bibfield  {title} {\bibinfo {title} {Striped superconductors: how spin,
  charge and superconducting orders intertwine in the cuprates},\ }\href
  {https://doi.org/10.1088/1367-2630/11/11/115004} {\bibfield  {journal}
  {\bibinfo  {journal} {New Journal of Physics}\ }\textbf {\bibinfo {volume}
  {11}},\ \bibinfo {pages} {115004} (\bibinfo {year} {2009})}\BibitemShut
  {NoStop}%
\bibitem [{\citenamefont {Vojta}(2009)}]{vojta09}%
  \BibitemOpen
  \bibfield  {author} {\bibinfo {author} {\bibfnamefont {M.}~\bibnamefont
  {Vojta}},\ }\bibfield  {title} {\bibinfo {title} {Lattice symmetry breaking
  in cuprate superconductors: stripes, nematics, and superconductivity},\
  }\href {https://doi.org/10.1080/00018730903122242} {\bibfield  {journal}
  {\bibinfo  {journal} {Advances in Physics}\ }\textbf {\bibinfo {volume}
  {58}},\ \bibinfo {pages} {699} (\bibinfo {year} {2009})}\BibitemShut
  {NoStop}%
\bibitem [{\citenamefont {Huang}\ \emph {et~al.}(2018)\citenamefont {Huang},
  \citenamefont {Mendl}, \citenamefont {Jiang}, \citenamefont {Moritz},\ and\
  \citenamefont {Devereaux}}]{huang18}%
  \BibitemOpen
  \bibfield  {author} {\bibinfo {author} {\bibfnamefont {E.~W.}\ \bibnamefont
  {Huang}}, \bibinfo {author} {\bibfnamefont {C.~B.}\ \bibnamefont {Mendl}},
  \bibinfo {author} {\bibfnamefont {H.-C.}\ \bibnamefont {Jiang}}, \bibinfo
  {author} {\bibfnamefont {B.}~\bibnamefont {Moritz}},\ and\ \bibinfo {author}
  {\bibfnamefont {T.~P.}\ \bibnamefont {Devereaux}},\ }\bibfield  {title}
  {\bibinfo {title} {Stripe order from the perspective of the {H}ubbard
  model},\ }\href {https://doi.org/10.1038/s41535-018-0097-0} {\bibfield
  {journal} {\bibinfo  {journal} {npj Quantum Materials}\ }\textbf {\bibinfo
  {volume} {3}},\ \bibinfo {pages} {1} (\bibinfo {year} {2018})}\BibitemShut
  {NoStop}%
\bibitem [{\citenamefont {Agterberg}\ \emph {et~al.}(2020)\citenamefont
  {Agterberg}, \citenamefont {Davis}, \citenamefont {Edkins}, \citenamefont
  {Fradkin}, \citenamefont {Van~Harlingen}, \citenamefont {Kivelson},
  \citenamefont {Lee}, \citenamefont {Radzihovsky}, \citenamefont {Tranquada},\
  and\ \citenamefont {Wang}}]{agterberg20}%
  \BibitemOpen
  \bibfield  {author} {\bibinfo {author} {\bibfnamefont {D.~F.}\ \bibnamefont
  {Agterberg}}, \bibinfo {author} {\bibfnamefont {J.~S.}\ \bibnamefont
  {Davis}}, \bibinfo {author} {\bibfnamefont {S.~D.}\ \bibnamefont {Edkins}},
  \bibinfo {author} {\bibfnamefont {E.}~\bibnamefont {Fradkin}}, \bibinfo
  {author} {\bibfnamefont {D.~J.}\ \bibnamefont {Van~Harlingen}}, \bibinfo
  {author} {\bibfnamefont {S.~A.}\ \bibnamefont {Kivelson}}, \bibinfo {author}
  {\bibfnamefont {P.~A.}\ \bibnamefont {Lee}}, \bibinfo {author} {\bibfnamefont
  {L.}~\bibnamefont {Radzihovsky}}, \bibinfo {author} {\bibfnamefont {J.~M.}\
  \bibnamefont {Tranquada}},\ and\ \bibinfo {author} {\bibfnamefont
  {Y.}~\bibnamefont {Wang}},\ }\bibfield  {title} {\bibinfo {title} {The
  physics of pair-density waves: cuprate superconductors and beyond},\ }\href
  {https://doi.org/10.1146/annurev-conmatphys-031119-050711} {\bibfield
  {journal} {\bibinfo  {journal} {Annual Review of Condensed Matter Physics}\
  }\textbf {\bibinfo {volume} {11}},\ \bibinfo {pages} {231} (\bibinfo {year}
  {2020})}\BibitemShut {NoStop}%
\bibitem [{\citenamefont {Tranquada}(2020)}]{tranquada20}%
  \BibitemOpen
  \bibfield  {author} {\bibinfo {author} {\bibfnamefont {J.~M.}\ \bibnamefont
  {Tranquada}},\ }\bibfield  {title} {\bibinfo {title} {Cuprate superconductors
  as viewed through a striped lens},\ }\href
  {https://doi.org/10.1080/00018732.2021.1935698} {\bibfield  {journal}
  {\bibinfo  {journal} {Advances in Physics}\ }\textbf {\bibinfo {volume}
  {69}},\ \bibinfo {pages} {437} (\bibinfo {year} {2020})}\BibitemShut
  {NoStop}%
\bibitem [{\citenamefont {Dehollain}\ \emph {et~al.}(2020)\citenamefont
  {Dehollain}, \citenamefont {Mukhopadhyay}, \citenamefont {Michal},
  \citenamefont {Wang}, \citenamefont {Wunsch}, \citenamefont {Reichl},
  \citenamefont {Wegscheider}, \citenamefont {Rudner}, \citenamefont {Demler},\
  and\ \citenamefont {Vandersypen}}]{Dehollain2020}%
  \BibitemOpen
  \bibfield  {author} {\bibinfo {author} {\bibfnamefont {J.~P.}\ \bibnamefont
  {Dehollain}}, \bibinfo {author} {\bibfnamefont {U.}~\bibnamefont
  {Mukhopadhyay}}, \bibinfo {author} {\bibfnamefont {V.~P.}\ \bibnamefont
  {Michal}}, \bibinfo {author} {\bibfnamefont {Y.}~\bibnamefont {Wang}},
  \bibinfo {author} {\bibfnamefont {B.}~\bibnamefont {Wunsch}}, \bibinfo
  {author} {\bibfnamefont {C.}~\bibnamefont {Reichl}}, \bibinfo {author}
  {\bibfnamefont {W.}~\bibnamefont {Wegscheider}}, \bibinfo {author}
  {\bibfnamefont {M.~S.}\ \bibnamefont {Rudner}}, \bibinfo {author}
  {\bibfnamefont {E.}~\bibnamefont {Demler}},\ and\ \bibinfo {author}
  {\bibfnamefont {L.~M.~K.}\ \bibnamefont {Vandersypen}},\ }\bibfield  {title}
  {\bibinfo {title} {Nagaoka ferromagnetism observed in a quantum dot
  plaquette},\ }\href {https://doi.org/10.1038/s41586-020-2051-0} {\bibfield
  {journal} {\bibinfo  {journal} {Nature}\ }\textbf {\bibinfo {volume} {579}},\
  \bibinfo {pages} {528} (\bibinfo {year} {2020})}\BibitemShut {NoStop}%
\bibitem [{\citenamefont {Manmana}\ \emph {et~al.}(2005)\citenamefont
  {Manmana}, \citenamefont {Muramatsu},\ and\ \citenamefont
  {Noack}}]{Manmana2005}%
  \BibitemOpen
  \bibfield  {author} {\bibinfo {author} {\bibfnamefont {S.~R.}\ \bibnamefont
  {Manmana}}, \bibinfo {author} {\bibfnamefont {A.}~\bibnamefont {Muramatsu}},\
  and\ \bibinfo {author} {\bibfnamefont {R.~M.}\ \bibnamefont {Noack}},\
  }\bibfield  {title} {\bibinfo {title} {Time evolution of one-dimensional
  quantum many body systems},\ }\href {https://doi.org/10.1063/1.2080353}
  {\bibfield  {journal} {\bibinfo  {journal} {AIP Conf. Proc.}\ }\textbf
  {\bibinfo {volume} {789}},\ \bibinfo {pages} {269} (\bibinfo {year}
  {2005})}\BibitemShut {NoStop}%
\bibitem [{\citenamefont {Prelov{\v{s}}ek}\ and\ \citenamefont
  {Bon{\v{c}}a}(2013)}]{Prelovsek}%
  \BibitemOpen
  \bibfield  {author} {\bibinfo {author} {\bibfnamefont {P.}~\bibnamefont
  {Prelov{\v{s}}ek}}\ and\ \bibinfo {author} {\bibfnamefont {J.}~\bibnamefont
  {Bon{\v{c}}a}},\ }\bibinfo {title} {Ground state and finite temperature
  {L}anczos methods},\ in\ \href {https://doi.org/10.1007/978-3-642-35106-8_1}
  {\emph {\bibinfo {booktitle} {Strongly Correlated Systems: Numerical
  Methods}}},\ \bibinfo {editor} {edited by\ \bibinfo {editor} {\bibfnamefont
  {A.}~\bibnamefont {Avella}}\ and\ \bibinfo {editor} {\bibfnamefont
  {F.}~\bibnamefont {Mancini}}}\ (\bibinfo  {publisher} {Springer Berlin
  Heidelberg},\ \bibinfo {address} {Berlin, Heidelberg},\ \bibinfo {year}
  {2013})\ pp.\ \bibinfo {pages} {1--30}\BibitemShut {NoStop}%
\bibitem [{\citenamefont {Zheng}\ \emph {et~al.}(2017)\citenamefont {Zheng},
  \citenamefont {Chung}, \citenamefont {Corboz}, \citenamefont {Ehlers},
  \citenamefont {Qin}, \citenamefont {Noack}, \citenamefont {Shi},
  \citenamefont {White}, \citenamefont {Zhang},\ and\ \citenamefont
  {Chan}}]{Zheng2017}%
  \BibitemOpen
  \bibfield  {author} {\bibinfo {author} {\bibfnamefont {B.-X.}\ \bibnamefont
  {Zheng}}, \bibinfo {author} {\bibfnamefont {C.-M.}\ \bibnamefont {Chung}},
  \bibinfo {author} {\bibfnamefont {P.}~\bibnamefont {Corboz}}, \bibinfo
  {author} {\bibfnamefont {G.}~\bibnamefont {Ehlers}}, \bibinfo {author}
  {\bibfnamefont {M.-P.}\ \bibnamefont {Qin}}, \bibinfo {author} {\bibfnamefont
  {R.~M.}\ \bibnamefont {Noack}}, \bibinfo {author} {\bibfnamefont
  {H.}~\bibnamefont {Shi}}, \bibinfo {author} {\bibfnamefont {S.~R.}\
  \bibnamefont {White}}, \bibinfo {author} {\bibfnamefont {S.}~\bibnamefont
  {Zhang}},\ and\ \bibinfo {author} {\bibfnamefont {G.~K.-L.}\ \bibnamefont
  {Chan}},\ }\bibfield  {title} {\bibinfo {title} {Stripe order in the
  underdoped region of the two-dimensional {H}ubbard model},\ }\href
  {https://doi.org/10.1126/science.aam7127} {\bibfield  {journal} {\bibinfo
  {journal} {Science}\ }\textbf {\bibinfo {volume} {358}},\ \bibinfo {pages}
  {1155} (\bibinfo {year} {2017})}\BibitemShut {NoStop}%
\bibitem [{\citenamefont {Huang}\ \emph {et~al.}(2017)\citenamefont {Huang},
  \citenamefont {Mendl}, \citenamefont {Liu}, \citenamefont {Johnston},
  \citenamefont {Jiang}, \citenamefont {Moritz},\ and\ \citenamefont
  {Devereaux}}]{Huang2017}%
  \BibitemOpen
  \bibfield  {author} {\bibinfo {author} {\bibfnamefont {E.~W.}\ \bibnamefont
  {Huang}}, \bibinfo {author} {\bibfnamefont {C.~B.}\ \bibnamefont {Mendl}},
  \bibinfo {author} {\bibfnamefont {S.}~\bibnamefont {Liu}}, \bibinfo {author}
  {\bibfnamefont {S.}~\bibnamefont {Johnston}}, \bibinfo {author}
  {\bibfnamefont {H.-C.}\ \bibnamefont {Jiang}}, \bibinfo {author}
  {\bibfnamefont {B.}~\bibnamefont {Moritz}},\ and\ \bibinfo {author}
  {\bibfnamefont {T.~P.}\ \bibnamefont {Devereaux}},\ }\bibfield  {title}
  {\bibinfo {title} {Numerical evidence of fluctuating stripes in the normal
  state of high-${T}_c$ cuprate superconductors},\ }\href
  {https://doi.org/10.1126/science.aak9546} {\bibfield  {journal} {\bibinfo
  {journal} {Science}\ }\textbf {\bibinfo {volume} {358}},\ \bibinfo {pages}
  {1161} (\bibinfo {year} {2017})}\BibitemShut {NoStop}%
\bibitem [{\citenamefont {Ponsioen}\ \emph {et~al.}(2019)\citenamefont
  {Ponsioen}, \citenamefont {Chung},\ and\ \citenamefont
  {Corboz}}]{Ponsioen19}%
  \BibitemOpen
  \bibfield  {author} {\bibinfo {author} {\bibfnamefont {B.}~\bibnamefont
  {Ponsioen}}, \bibinfo {author} {\bibfnamefont {S.~S.}\ \bibnamefont
  {Chung}},\ and\ \bibinfo {author} {\bibfnamefont {P.}~\bibnamefont
  {Corboz}},\ }\bibfield  {title} {\bibinfo {title} {Period 4 stripe in the
  extended two-dimensional {H}ubbard model},\ }\href
  {https://doi.org/10.1103/PhysRevB.100.195141} {\bibfield  {journal} {\bibinfo
   {journal} {Phys. Rev. B}\ }\textbf {\bibinfo {volume} {100}},\ \bibinfo
  {pages} {195141} (\bibinfo {year} {2019})}\BibitemShut {NoStop}%
\bibitem [{\citenamefont {D'Alessio}\ and\ \citenamefont
  {Rigol}(2014)}]{Alessio14}%
  \BibitemOpen
  \bibfield  {author} {\bibinfo {author} {\bibfnamefont {L.}~\bibnamefont
  {D'Alessio}}\ and\ \bibinfo {author} {\bibfnamefont {M.}~\bibnamefont
  {Rigol}},\ }\bibfield  {title} {\bibinfo {title} {Long-time behavior of
  isolated periodically driven interacting lattice systems},\ }\href
  {https://doi.org/10.1103/PhysRevX.4.041048} {\bibfield  {journal} {\bibinfo
  {journal} {Phys. Rev. X}\ }\textbf {\bibinfo {volume} {4}},\ \bibinfo {pages}
  {041048} (\bibinfo {year} {2014})}\BibitemShut {NoStop}%
\bibitem [{\citenamefont {Sorg}\ \emph {et~al.}(2014)\citenamefont {Sorg},
  \citenamefont {Vidmar}, \citenamefont {Pollet},\ and\ \citenamefont
  {Heidrich-Meisner}}]{Sorg14}%
  \BibitemOpen
  \bibfield  {author} {\bibinfo {author} {\bibfnamefont {S.}~\bibnamefont
  {Sorg}}, \bibinfo {author} {\bibfnamefont {L.}~\bibnamefont {Vidmar}},
  \bibinfo {author} {\bibfnamefont {L.}~\bibnamefont {Pollet}},\ and\ \bibinfo
  {author} {\bibfnamefont {F.}~\bibnamefont {Heidrich-Meisner}},\ }\bibfield
  {title} {\bibinfo {title} {Relaxation and thermalization in the
  one-dimensional {B}ose-{H}ubbard model: {A} case study for the interaction
  quantum quench from the atomic limit},\ }\href
  {https://doi.org/10.1103/PhysRevA.90.033606} {\bibfield  {journal} {\bibinfo
  {journal} {Phys. Rev. A}\ }\textbf {\bibinfo {volume} {90}},\ \bibinfo
  {pages} {033606} (\bibinfo {year} {2014})}\BibitemShut {NoStop}%
\bibitem [{\citenamefont {Bauer}\ \emph {et~al.}(2015)\citenamefont {Bauer},
  \citenamefont {Dorfner},\ and\ \citenamefont {Heidrich-Meisner}}]{Bauer15}%
  \BibitemOpen
  \bibfield  {author} {\bibinfo {author} {\bibfnamefont {A.}~\bibnamefont
  {Bauer}}, \bibinfo {author} {\bibfnamefont {F.}~\bibnamefont {Dorfner}},\
  and\ \bibinfo {author} {\bibfnamefont {F.}~\bibnamefont {Heidrich-Meisner}},\
  }\bibfield  {title} {\bibinfo {title} {Temporal decay of {N}\'eel order in
  the one-dimensional {F}ermi-{H}ubbard model},\ }\href
  {https://doi.org/10.1103/PhysRevA.91.053628} {\bibfield  {journal} {\bibinfo
  {journal} {Phys. Rev. A}\ }\textbf {\bibinfo {volume} {91}},\ \bibinfo
  {pages} {053628} (\bibinfo {year} {2015})}\BibitemShut {NoStop}%
\bibitem [{\citenamefont {Maier}\ \emph {et~al.}(2005)\citenamefont {Maier},
  \citenamefont {Jarrell}, \citenamefont {Schulthess}, \citenamefont {Kent},\
  and\ \citenamefont {White}}]{Maier2005}%
  \BibitemOpen
  \bibfield  {author} {\bibinfo {author} {\bibfnamefont {T.~A.}\ \bibnamefont
  {Maier}}, \bibinfo {author} {\bibfnamefont {M.}~\bibnamefont {Jarrell}},
  \bibinfo {author} {\bibfnamefont {T.~C.}\ \bibnamefont {Schulthess}},
  \bibinfo {author} {\bibfnamefont {P.~R.~C.}\ \bibnamefont {Kent}},\ and\
  \bibinfo {author} {\bibfnamefont {J.~B.}\ \bibnamefont {White}},\ }\bibfield
  {title} {\bibinfo {title} {Systematic study of $d$-wave superconductivity in
  the 2d repulsive {H}ubbard model},\ }\href
  {https://doi.org/10.1103/PhysRevLett.95.237001} {\bibfield  {journal}
  {\bibinfo  {journal} {Phys. Rev. Lett.}\ }\textbf {\bibinfo {volume} {95}},\
  \bibinfo {pages} {237001} (\bibinfo {year} {2005})}\BibitemShut {NoStop}%
\bibitem [{\citenamefont {F\"orst}\ \emph {et~al.}(2014)\citenamefont
  {F\"orst}, \citenamefont {Tobey}, \citenamefont {Bromberger}, \citenamefont
  {Wilkins}, \citenamefont {Khanna}, \citenamefont {Caviglia}, \citenamefont
  {Chuang}, \citenamefont {Lee}, \citenamefont {Schlotter}, \citenamefont
  {Turner}, \citenamefont {Minitti}, \citenamefont {Krupin}, \citenamefont
  {Xu}, \citenamefont {Wen}, \citenamefont {Gu}, \citenamefont {Dhesi},
  \citenamefont {Cavalleri},\ and\ \citenamefont {Hill}}]{Forst2014}%
  \BibitemOpen
  \bibfield  {author} {\bibinfo {author} {\bibfnamefont {M.}~\bibnamefont
  {F\"orst}}, \bibinfo {author} {\bibfnamefont {R.~I.}\ \bibnamefont {Tobey}},
  \bibinfo {author} {\bibfnamefont {H.}~\bibnamefont {Bromberger}}, \bibinfo
  {author} {\bibfnamefont {S.~B.}\ \bibnamefont {Wilkins}}, \bibinfo {author}
  {\bibfnamefont {V.}~\bibnamefont {Khanna}}, \bibinfo {author} {\bibfnamefont
  {A.~D.}\ \bibnamefont {Caviglia}}, \bibinfo {author} {\bibfnamefont {Y.-D.}\
  \bibnamefont {Chuang}}, \bibinfo {author} {\bibfnamefont {W.~S.}\
  \bibnamefont {Lee}}, \bibinfo {author} {\bibfnamefont {W.~F.}\ \bibnamefont
  {Schlotter}}, \bibinfo {author} {\bibfnamefont {J.~J.}\ \bibnamefont
  {Turner}}, \bibinfo {author} {\bibfnamefont {M.~P.}\ \bibnamefont {Minitti}},
  \bibinfo {author} {\bibfnamefont {O.}~\bibnamefont {Krupin}}, \bibinfo
  {author} {\bibfnamefont {Z.~J.}\ \bibnamefont {Xu}}, \bibinfo {author}
  {\bibfnamefont {J.~S.}\ \bibnamefont {Wen}}, \bibinfo {author} {\bibfnamefont
  {G.~D.}\ \bibnamefont {Gu}}, \bibinfo {author} {\bibfnamefont {S.~S.}\
  \bibnamefont {Dhesi}}, \bibinfo {author} {\bibfnamefont {A.}~\bibnamefont
  {Cavalleri}},\ and\ \bibinfo {author} {\bibfnamefont {J.~P.}\ \bibnamefont
  {Hill}},\ }\bibfield  {title} {\bibinfo {title} {Melting of charge stripes in
  vibrationally driven {L}a$_{1.875}${B}a$_{0.125}${C}u{O}$_{4}$: Assessing the
  respective roles of electronic and lattice order in frustrated
  superconductors},\ }\href {https://doi.org/10.1103/PhysRevLett.112.157002}
  {\bibfield  {journal} {\bibinfo  {journal} {Phys. Rev. Lett.}\ }\textbf
  {\bibinfo {volume} {112}},\ \bibinfo {pages} {157002} (\bibinfo {year}
  {2014})}\BibitemShut {NoStop}%
\bibitem [{\citenamefont {Wang}\ \emph
  {et~al.}(2021{\natexlab{b}})\citenamefont {Wang}, \citenamefont {Chen},
  \citenamefont {Devereaux}, \citenamefont {Moritz},\ and\ \citenamefont
  {Mitrano}}]{wang2021b}%
  \BibitemOpen
  \bibfield  {author} {\bibinfo {author} {\bibfnamefont {Y.}~\bibnamefont
  {Wang}}, \bibinfo {author} {\bibfnamefont {Y.}~\bibnamefont {Chen}}, \bibinfo
  {author} {\bibfnamefont {T.~P.}\ \bibnamefont {Devereaux}}, \bibinfo {author}
  {\bibfnamefont {B.}~\bibnamefont {Moritz}},\ and\ \bibinfo {author}
  {\bibfnamefont {M.}~\bibnamefont {Mitrano}},\ }\bibfield  {title} {\bibinfo
  {title} {X-ray scattering from light-driven spin fluctuations in a doped
  {M}ott insulator},\ }\href {https://doi.org/10.1038/s42005-021-00715-z}
  {\bibfield  {journal} {\bibinfo  {journal} {Communications Physics}\ }\textbf
  {\bibinfo {volume} {4}},\ \bibinfo {pages} {212} (\bibinfo {year}
  {2021}{\natexlab{b}})}\BibitemShut {NoStop}%
\bibitem [{\citenamefont {Chen}\ \emph {et~al.}(2010)\citenamefont {Chen},
  \citenamefont {Gu},\ and\ \citenamefont {Wen}}]{Chen2010}%
  \BibitemOpen
  \bibfield  {author} {\bibinfo {author} {\bibfnamefont {X.}~\bibnamefont
  {Chen}}, \bibinfo {author} {\bibfnamefont {Z.-C.}\ \bibnamefont {Gu}},\ and\
  \bibinfo {author} {\bibfnamefont {X.-G.}\ \bibnamefont {Wen}},\ }\bibfield
  {title} {\bibinfo {title} {Local unitary transformation, long-range quantum
  entanglement, wave function renormalization, and topological order},\ }\href
  {https://doi.org/10.1103/PhysRevB.82.155138} {\bibfield  {journal} {\bibinfo
  {journal} {Phys. Rev. B}\ }\textbf {\bibinfo {volume} {82}},\ \bibinfo
  {pages} {155138} (\bibinfo {year} {2010})}\BibitemShut {NoStop}%
\end{thebibliography}%

\end{document}